\documentclass[12pt]{iopart}

\usepackage{graphicx,color} 
\usepackage{bm}       
\usepackage{iopams} 
\usepackage[dvipsnames]{xcolor}
\usepackage{ulem}
\usepackage{siunitx}
\usepackage{multirow}
\usepackage{graphicx}
\begin{document}
\bibliographystyle{iopart-num}
\newcommand {\nc} {\newcommand}
\newcommand{\vv}[1]{{$\bf {#1}$}}
\newcommand{\vvm}[1]{{\bf {#1}}}
\def\btau{\mbox{\boldmath$\tau$}}
\nc{\half}{\frac{1}{2}}
\nc{\numberthis}{\addtocounter{equation}{1}\tag{\theequation}}
\nc{\lla}{\left\langle}
\nc{\rra}{\right\rangle}
\nc{\lrme}{\left|\left|}
\nc{\rrme}{\right|\right|}

\title{Adding Corrections to Global Spherical Potentials for Use in a Coupled-Channel Formulation}

\author{S.~P.~Weppner$^1$, Prakrut ~Patel$^1$, C.~Miller$^1$, C.~Ogg$^1$, and I.~Mazumdar$^2$}
\address{ $^1$ Natural Sciences, Eckerd College, St. Petersburg, FL 33711, USA}
\ead{weppnesp@eckerd.edu}

\address{$^2$ Tata Institute of Fundamental Research, 400005 Mumbai, India}

\date{August  2023}

\begin{abstract}
	The coupled-channel technique augments
	a non-relativistic distorted wave born approximation scattering calculation to include a coupling to virtual states from the negative
		energy region. It has been found to be important in low energy nucleon-nucleus scattering.
We modify the nucleon-nucleus standard optical potentials, not designed for a coupled-channel space, so
	they can be used in that setting. The changes are small and systematic.
We use a standard scattering code to adjust a variety of optical potentials and targets
	such that the original fit to scattering
observables are maintained as we incorporate the coupled-channel environment. Overall over forty
target nuclei were tested from $A=12$ to $A=205$ and nucleon projectile energies from 1 MeV to 200 MeV.
There is excellent improvement in fitting the scattering observables, especially
	for low energy neutron scattering.
                The corrections were found to be unimportant for projectile
                energies greater than 200 MeV. 
		The largest changes are to the surface amplitudes while the real radii and the real central amplitude are 
		modified by only a few percent, every other parameter is unchanged.
 This technique is general enough to be applied to a variety of inelastic theoretical calculations.
\end{abstract}

\maketitle

\section{Introduction}
\label{sec:intro}

The optical model nucleon-nucleus potential~\cite{https://doi.org/10.48550/arxiv.2210.07293, DICKHOFF2019252} 
within a coupled-channel (CC) approach has
a rich history (see an excellent review by Tamura~\cite{doi:10.1146/annurev.ns.19.120169.000531}). 
The coupled-channel approach augments a non-relativistic Schr\"{o}dinger approach by coupling the
positive energy scattering problem with negative energy 
resonances, bound states, spectroscopic factors and transition potentials~\cite{PhysRevC.71.057602,Thompson:1988zz}. 
A common issue is that
the spherical distorted wave Born approximation (DWBA) optical potential is often phenomenologically fit lacking the coupled-channel context.
The most widely used spherical global phenomenological optical model to date is from Koning and
Delaroche~\cite{KONING2003231} but there have been plenty of other 
similar attempts~\cite{PhysRevC.80.034608,globalM,globalB,globalJ,Ca40p80ra,globalV,globalR,globalMR,globalC, PhysRevLett.127.182502, PhysRevC.107.014602}. 
These models, 
being created in a DWBA context, are not directly suited 
for CC calculations especially at low energies~\cite{Nobre_2015}. There have been some attempts
to fit the optical potential directly 
in a coupled-channel environment~\cite{EfremShSoukhovitskii_2004, doi:10.1080/18811248.2008.9711442,PhysRevC.94.064605, HAGINO2022103951}.
There also has been work on adjusting the scattering (S) matrix of a single reaction DWBA optical potential to make it suitable for the
coupled-channel calculation~\cite{PhysRevC.104.044616}.
The present work takes another approach
which is to augment the DWBA potential parameters in a global fashion by making slight modifications so that
they are suitable for a coupled-channel methodology as other researchers have attempted in limited 
forms~\cite{1980ZPhyA297223B,talys_web,Nobre_2015, 2018EPJA,PhysRevC.93.064311}.
In the present work we expand these previous efforts by generalizing the corrections for the spherical DWBA 
to make these potentials appropriate for coupled-channel calculations. The improvements introduced are 
valid for a large range of projectile energies,
target nuclei, and potentials.

To be clear, at low energy a spherical DWBA optical potential is not sufficient and other augmentations to the theory are needed including
channel coupling, compound nucleus formation, and spectroscopic factors. However the researcher still needs functional optical potentials for the 
reactions and most often the spherical DWBA optical potential
is the only available choice. If this spherical DWBA  optical potential fits the elastic scattering data to a high degree and also has been fit to the total reaction experimental data and
one wants these fits to be maintained than this prescription is a viable easy choice.
A word on the importance of why the standard DWBA optical potentials must be changed before they are 
considered viable for CC calculations. 
The optical potential theory treats the target nucleus like a cloudy glass sphere where the elastic observables are treated 
directly (refracting from the sphere with no parameters) and the inelastic is described in only a gross fashion as
the projectile beam that is absorbed by the cloudy sphere (with only the total 
inelastic cross-sections produced) and thus lost to other channels. 
Over the years methods using this optical potential have been devised to calculate specific inelastic 
reactions: excitation, breakup, exchange, fission, etc. These inelastic methods will include other parameters, 
transition potentials, or spectroscopic factors which are not included in the original optical potential. 
For the potential to remain consistent  when the  Hilbert space is changed to a non-spherical coupled channel 
we must maintain the integrity of the original refractive and absorptive properties. A good optical potential had these properties 
fit originally to experimental values (using elastic differential cross-sections, elastic spin experiments, 
and total reaction cross-sections) and thus it is essential that the integrity of these observables be 
maintained before any additional inelastic ingredients are included which require these equivalent gross 
properties for consistency.


\section{Theoretical Frameworks}
\label{sec:theory}

The phenomenological optical model potentials~\cite{KONING2003231,PhysRevC.80.034608,PhysRevC.89.049904} used in this work 
contain the traditional
volume ($V$), surface ($S$), and spin-orbit ($SO$) nuclear terms which are
delineated using the
standard Woods-Saxon form factors
\begin{equation}
	f_{WS}(r,{\cal R}_i,{\cal A}_i)=(1+\exp\left ((r-{\cal R}_i)/{\cal A}_i)\right)^{-1}, \label{ws1}
\end{equation}
where ${\cal R}_i$ is the radius parameter and ${\cal A}_i$ is the diffusive parameter.
The $i$ is a placeholder for the $V$, $S$, or $SO$ designation.
The phenomenological optical model potentials takes a standard form:
\begin{eqnarray}
&&{\cal U}(r,E,A,N,Z)= \nonumber \\
&&\Big(-{\cal V}_V(E,A,N,Z)-i{\cal W}_V(E,A,N,Z)\Big)
f_{WS}(r,{\cal R}_V,{\cal A}_V) \nonumber \\
&&+4{\cal A}_S\Big({\cal V}_D(E,A)+i{\cal W}_D(E,A,N,Z)\Big)
\frac{d}{dr}f_{WS}(r,{\cal R}_D,{\cal A}_D) \nonumber \\
&&+\frac{2}{r}\Big({\cal V}_{SO}(E,A,N,Z)+i{\cal W}_{SO}(E,A,N,Z)\Big)
\frac{d}{dr}f_{WS}(r,{\cal R}_{SO},{\cal A}_{SO})({\bf l\cdot\sigma}) \nonumber \\
&&+f_{coul}(r,{\cal R}_C,A,N,Z)
\label{WS}
\end{eqnarray}
where the ${\cal V}_i$ and ${\cal W}_i$ are the real and imaginary potential
amplitudes respectively and $f_{coul}(r,{\cal R}_C,A,N,Z)$ is the
coulomb term which has the following traditional format with a proton
projectile:
\begin{eqnarray}
f_{coul}(r,{\cal R}_C,A,N,Z) = \frac{Ze^2}{r},\;\;\;\;\;r\ge {\cal R}_C,
\nonumber \\
f_{coul}(r,{\cal R}_C,A,N,Z) = \frac{Ze^2}{2{\cal R}_C}
\Big(3-\frac{r^2}{{{\cal R}_C}^2}\Big ), r\le {\cal R}_C. \label{coulomb}
\end{eqnarray}
For a neutron projectile this term is set to zero.
There were two optical potentials used in this work. Mainly we used the aforementioned Koning and
Delaroche~\cite{KONING2003231} which covers a large range of energies and targets. For targets
below  an atomic number (A) of 27 we used the 
potential of one of the authors~\cite{PhysRevC.80.034608, PhysRevC.89.049904}.
We believe that using two different potentials has increased the robustness of this technique.

In earlier research optical potentials have been modified 
so that they function in a coupled-channel (CC) approach 
so we give a brief overview
of these past efforts. The {\it TALYS} scattering code~\cite{talys,talys_web}, 
which has as its direct scattering tool  Raynal's {\it ECIS}~\cite{PhysRevC.71.057602,Raynal:94,ECIS06}, 
state
in the manual (all versions up through the present) that if a CC calculation is attempted 
and no potential is
given then {\it TALYS} will use the DWBA potential KD03~\cite{KONING2003231} (our choice for $A>26$)
but will reduce the imaginary surface term by 15\% (which
is denoted as ${\cal W}_D(E,A)$ in Eq.~\ref{WS} of this work).  Later Nobre et. al~\cite{Nobre_2015} used a suggestion by 
Bang and Vaagen~\cite{1980ZPhyA297223B}
who were interested in studying the effects of deformation in nuclear orbitals. To conserve volume they modified the radius terms of the central
potential, denoted above  as ${\cal R}_V$ (volume) and ${\cal R}_D$ (surface)
\begin{equation}
	{{\cal R^\prime}_{i=V,D}} = {{\cal R}_i}\big(1-\frac{{\sum_\lambda}\beta_\lambda^2}{4\pi}\big ) \label{root_eq},
\end{equation}	
where $\beta_\lambda$ are the deformation parameters which make the nucleus non-spherical:
\begin{equation}
	{\cal R}^\prime_i(\theta,\phi) = R_i(1 + \sum_\lambda \beta_\lambda Y^0_\lambda).
\end{equation}
These researchers, using the modification prescribed in Eq.~\ref{root_eq}, 
found satisfactory results for neutron scattering off the rare-earth nuclei. 
Al-Rawashdeh and Jaghoub continued this investigation with the same modification to the
actinide nuclei (A=227-250) and came to similar conclusions~\cite{2018EPJA}. 

The corrections mentioned were relatively small~\cite{talys_web,Nobre_2015, 2018EPJA}. For the present research we started with
these same previous modifications: we lower the imaginary surface term, we also adjust the radius for the real volume, surface 
{\bf and additionally the spin-orbit terms} as 
prescribed. Instead of a static 15\% drop of the imaginary surface term 
we fit the imaginary and real surface terms to the original
elastic and total reaction observables.
We also slightly adjusted the depth of the real part of the central amplitude in an analytical fashion. 
This entire process was done for a large range of mass numbers (A=12-205) for the target
and energies (1 MeV-205 MeV) for either the neutron and proton projectile. 
Many of the optical potential parameters were not adjusted hoping that this retains the original flavor of the DWBA potential
(all diffusive parameters, ${\cal A_i}$, all imaginary radii, Im ${\cal R_i}$, 
the spin-orbit amplitude parameters $({\cal V}_{SO}, {\cal W}_{SO})$, and the coulomb parameter (${\cal R}_C$) 
were untouched). 

The most complicated aspect of this alteration was 
the modification of  the surface which we believe is the aspect of the nucleus most changed
by the distortion. Many researches have studied the importance of the surface term, and its corresponding surface energy,  in nuclear 
reactions~\cite{DANIELEWICZ2003233, hasse2012geometrical, PhysRevC.85.024606, PhysRevC.91.064308, PhysRevC.89.024619,Weppner_2018}. 
The skin thickness of the
nuclear matter plays a large roll in mimicking a 
distortion effect important in a coupled-channel calculation.
Finding a standard prescription for the surface term was a struggle, the end result was
to rely only on the observables and treat each set of reactions (fixed energy of projectile and target nucleus)
as an independent entity
thus fitting each surface amplitude to the entire set of results (cross-sections and spin observables).
We had these realistic constraints: \\
$\bullet$ All elastic measurements and all total reaction cross-sections should be as closely preserved as 
possible to the original potential when it is used in a coupled-channel environment. \\
$\bullet$ As $\frac{{\sum_\lambda}\beta_\lambda^2}{4\pi} \to 0$ then the distortion effect should also head to zero. \\
$\bullet$ As the target size, $A$, approaches 200 the distortion dependence on $A$ should saturate \\
$\bullet$ If the projectile energy of the nucleon is near zero the distortion effect should be at a maximum. \\
$\bullet$ As the projectile energy of the nucleon increases the distortion effect should diminish towards zero. \\
$\bullet$ The surface terms should be modified with the ansatz: the smaller the change the better.\\
Although fitting the surface term is not analytical it provides the flexability needed for adjustments made to the
theoretical system (number of channels and strength of each channel).

The total cross-sections deserve special mention in the thought process of our developed theory since
they represent an interplay between the strength of the real and imaginary scattering amplitudes of the 
solution to the DWBA scattering problem. Since Eq.~\ref{root_eq} reduces the radius this will affect the strength
(the volume) of the reaction. This equation, as derived by Bang and Vaagen, was the first order term in 
an expansion to conserve the volume of the central term of the Woods-Saxon~\ref{ws1} during distortion. 
We agree that the volume should not change by much but the true
measure of the quality of a modification (ie keep the strength of the potential the same)
is to maintain the results of the proton-nucleus reaction cross-section and the neutron-nucleus reaction
and total cross-section. The analytical function  which approximates the solution to the  
volume integral of the Woods-Saxon, given in Ref.~\cite{KONING2003231}, as
\begin{eqnarray}
	{\cal J}_V &= \int_{all space}(1+\exp\left ((r-{\cal R}_V)/{\cal A}_V)\right)^{-1}\:d^3{\vec{r}} \nonumber \\
	&= \frac{4}{3}\pi{\cal R}_V^3 (1 + (\frac{{\pi\cal A}_V}{{\cal R}_V})^2)
\end{eqnarray}
is illustrative. We found, in the spirit of the phenomenological approach, that what worked
best in maintaining the total cross-section observables was to adjust the real central amplitude
by a prescription that uses the above equation. We found that the spherical volume
was nearly conserved as was the magnitude of the surface term but
ultimately our test was the measurable quantities (emphasizing not fitting to experimental results but instead
fitting the distorted CC theoretical calculation to the original DWBA non-CC theoretical 
calculation). Our principle measure was therefore
a weighted $\chi^2$ where we solely compared the difference in calculated theoretical observables. We also 
deemed the manner of the weighting of this $\chi^2$ to be important.
The most important observables
were the total neutron cross-sections, total reaction cross-sections, and the forward angles
of the differential cross-section (with a regressive sliding scale as the angle increases). 
Of lesser importance were the
backward angles of the differential cross-section, and the forward angles of the spin polarization, 
and of least importance were the backward angles of the polarization. For example the percent error 
for the differential cross-section standard deviation at 20, 90, and 160 degrees was  5\%, 18\%, and 31\% respectively. Note that since most quality
optical potentials are fit to a plethora of experimental data we felt that experiment was not being ignored,
ultimately maintaining  the quality of the original optical potential
fit to the experimental elastic data and
total inelastic data was the objective.

\begin{figure}\includegraphics[width=3.48in]{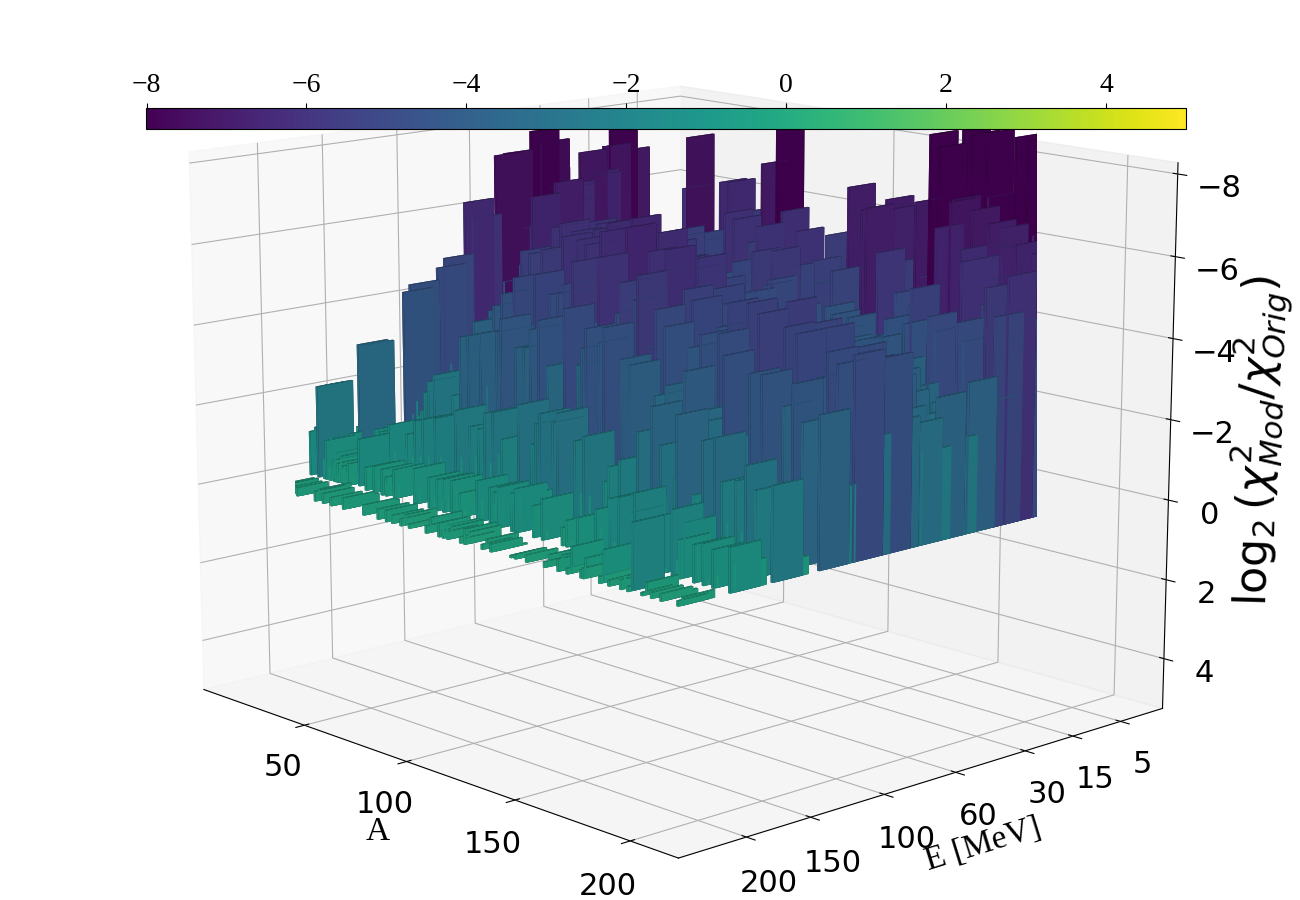}
	\caption{This shows the improvement on all of our results as a function of atomic number (A) and projectile
	energy (E). The amplitude is a ratio of the  $\chi^2$ on the scattering observables
	for our modified coupled channel potential and the original optical potential both used in a 
	coupled-channel environment. Note that this is on a $\log_2$ scale so a -8 refers to
	a reduction of a factor of 256 in the the $\chi^2$, more discussion is in the text.}
\end{figure}\label{fig1}

	Here follows a mathematical summary of our analytical changes which have been discussed in this section.
	First some definitions:
\begin{eqnarray}
	{{\cal R}_i}^\prime &=& (1 - \Delta_R) {\cal R}_i \nonumber \\
	\Delta_R &=& \frac{60^4}{(E^2+60^2)^2}\frac{{\sum_\lambda}\beta_\lambda^2}{4\pi} \nonumber \\
	\frac{{\cal J}_V}{{{\cal J}_V}^\prime} &=& (\frac{{\cal R}_V}{{{\cal R}_V}^\prime})^3
	\frac{1+(\pi a_V/{\cal R}_V)^2}{1+(\pi a_V/{{\cal R}_V}^\prime)^2} \label{DeltaR}.
\end{eqnarray}	
The volume integrals are ${{\cal J}_V}$, and we use the approximate analytical version found in Ref.~\cite{KONING2003231}.
The radius is changed by a $\Delta_R$ but there is a sliding energy scale attached to it, $\frac{60^4}{(E^2+60^2)^2}$,
this was not in earlier work~\cite{Nobre_2015, 2018EPJA} which had as a focus low energy scattering. We found
that the strongest dependence exhibited by coupled-channel calculations is projectile energy. At over 200 MeV 
the coupled-channel is nearly insignificant (see the next section for discussion) so we fitted an energy scale to our modifications. Now we list
the four analytical changes:
\begin{eqnarray}
	Re({{\cal R}_V}^\prime) &=& (1-\Delta_R) Re({{\cal R}_V}) \nonumber \\
	Re({{\cal R}_D}^\prime) &=& (1-\Delta_R) Re({{\cal R}_D}) \nonumber \\
	Re({{\cal R}_{SO}}^\prime) &=& (1-\Delta_R) Re({{\cal R}_{SO}}) \nonumber \\
	{{\cal V}_V}^\prime &=& (1-\Delta_R)^3\frac{{\cal J}_V}{{{\cal J}_V}^\prime}{{\cal V}_V}_0  \nonumber \\
			    &=& \frac{1+(\pi a_V/{\cal R}_V)^2}{1+(\pi a_V/{{\cal R}_V}^\prime)^2} {{\cal V}_V}_0.  
\end{eqnarray}
Lastly the optical potential
fitted the real and imaginary surface amplitudes directly to the observables, most modern
codes have this capability ({\it TALYS}~\cite{talys,talys_web}, 
{\it FRESCO}~\cite{FRESCO,Thompson:1988zz}, {\it ECIS}~\cite{PhysRevC.71.057602,Raynal:94,ECIS06}, 
and {\it EMPIRE}~\cite{Herman:2007}). At most the surface amplitudes were changed by 6 MeV (upward
for the real and downward for the imaginary). 
In summary a total of six parameters were changed to make these potentials CC ready. The surface terms were left adjustable 
because the coupling varies depending on the reaction or the theoretical model used. This prescription
dictates a fixed change to the real radii, the real central amplitude, and the real spin-orbit amplitude but the
reaction-model dependent fit is based on fitting the surface amplitudes correctly.


\section{Results and Discussion}a
\label{sec:results}
\subsection{General Results}

\begin{table}
\centering
 \begin{tabular}{||c ||c| c| c||}
 \hline
	 Energy  &p/n & $\chi^2_0$ & $\chi^2_f$ \\ 
	 \hline [MeV] &&&\\
 \hline\hline
	 \multirow{2}{*}{1.0} & n & $2.1\times10^4$ & $1.5\times10^2$ \\
			      & p &  $0.0$     &  $0.0$   \\ \hline
	 \multirow{2}{*} {5.0} & n & $4.2\times10^3$ & $1.3\times10^2$ \\
	                     & p &   $7.3\times10^1$    &  $1.1\times10^0$   \\ \hline
	 \multirow{2}{*}{10.0} & n & $1.7\times10^3$ & $5.6\times10^1$ \\
	                     & p &  $5.9\times10^2$     &  $2.3\times10^1$   \\ \hline
	 \multirow{2}{*}{ 20.0} &n&  $8.7\times10^2$   & $6.4\times10^1$      \\	 
	                     & p &  $1.7\times10^2$     &$2.0\times10^1$     \\ \hline
	 \multirow{2}{*}{ 35.0} & n & $9.3\times10^2$ & $5.1\times10^1$ \\
	                     & p &  $4.7\times10^2$     & $5.6\times10^1$    \\\hline
	 \multirow{2}{*}{ 65.0} & n & $6.3\times10^2$ & $2.7\times10^1$ \\
	                     & p &  $4.1\times10^2$     & $3.4\times10^1$    \\\hline
	 \multirow{2}{*}{100.0} & n & $2.0\times10^2$     &  $3.0\times10^1$    \\
	                     & p &  $2.5\times10^2$     & $4.3\times10^1$    \\\hline
	 \multirow{2}{*}{130.0} & n & $1.0\times10^2$   & $2.8\times10^1$    \\
	                     & p &    $7.6\times10^1$   & $4.0\times10^1$    \\\hline
	 \multirow{2}{*}{200.0} &  n  &$4.5\times10^1$  & $4.0\times10^1$    \\
	                     & p &    $4.4\times10^1$   & $3.6\times10^1$    \\
 \hline
 \end{tabular}
	\caption{This summarizes our results at nine different projectile energies separated by 
	the isospin of the projectile (p/n). The observables are a weighted chi squared which involved
	fitting to the differential cross-section, the differential polarization, the total
	reaction cross-section, and in the neutron projectile case the total cross-section.
	The initial chi squared ($\chi^2_0$) is the average $\chi^2$ difference between
	the unmodified optical potential in the couple channel calculation and the 
	calculation not including coupled-channels. The final chi-squared ($\chi^2_f$)
	is nearly the same calculation - except we have added the improvements to the optical
	potential when using
	the coupled-channel. These are statistical averages but each value is the average
	of at least 25 points,  for more analysis see the text.}
\label{T1}
\end{table}

Using the weighted $\chi^2$ ratio as our test, we tested over 40 nuclei at over 10 different energies and two projectiles (the proton and the
neutron), thus over 1000 cases. Each case computed the differential cross-section, the polarization, the total reaction cross-section, 
and for a neutron projectile the total cross-section. Each case was run as a normal DWBA (using {\it ECIS}), then we ran it again (the  
same unmodified optical potential) in a coupled channel framework (again using {\it ECIS}). 
In this instance the potential was used incorrectly; a global spherical optical potential 
used in a 
coupled channel space. The purpose was as the control to see how much the couple-channel space distorted the original observable results. 
Finally it was run with our
adjustments in this coupled-channel methodology
to ascertain if we could bring the calculated quantities back to their original values.

\begin{figure*}\includegraphics[width=6.95in]{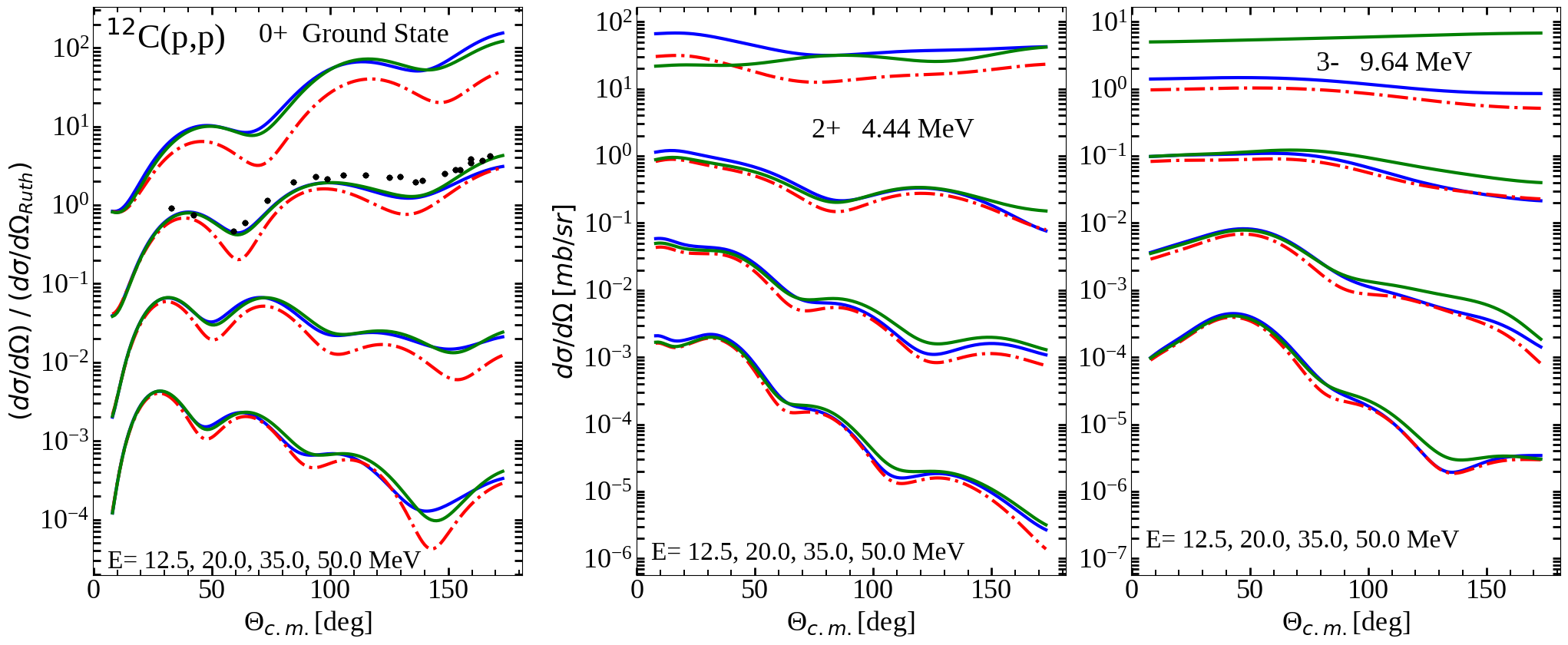}
	\caption{Differential cross-sections for a proton on $^{12}$C. In all calculations the solid blue line is a calculation using the original spherical
	DWBA potential, the solid green line has modified that potential and uses a coupled-channel calculation. The red dashed line is the
	inconsistent hybrid of the original potential used in a coupled-channel environment. The left panel represents the elastic (ground state only)
	differential cross-section divided by the Rutherford cross-Section
	for four different proton energies, lowest energy at the top, highest energy at the bottom. This left panel presents one of the observables
	which we fit to, a good fit would be when the solid green calculation is close to the original solid blue calculation. The central and right
	panel represent inelastic differential cross-sections which were not used in the fittings, they however show the sensitivity of the modification
	to the collective 2+ and 3- or 4+ inelastic cross-sections, again the projectile energies are the lowest at the top and highest at the bottom. For more
	analysis see the text.
	For aesthetics the  E=20.0 MeV cross-sections have been reduced by a factor of 20, the E= 35.0 MeV by a factor of 400, the E = 50.0 by a factor of 8000.
	The experimental data at E = 20.0 MeV is from Ref.~\cite{Tesmer:1972zz}. For $^{12}$C $\beta_2 = -0.71$, $\beta_3 = 0.42$.}
\label{figds}\end{figure*}

The results of our modifications
were very good, our weighted $\chi^2$ improvement ratio averaged 16 which our worst case was 1.25 (at high energy) and our best case
was 400 (see Fig.~1). That is the weighted $\chi^2$ was reduced, on average, by {\bf a factor of} 16. Figure~\ref{fig1} shows that the best improvement happened
systematically at low energies, where the influence of coupled channels is the greatest. The smallest improvement happened at high energies,
where the coupled channel effect is the least influential. This was our motivation for Eq.~\ref{DeltaR} where the
projectile energy scale factor $\frac{60^4}{(E^2+60^2)^2}$ quickly suppresses any changes such that at a projectile energy of 
240 MeV the suppression factor for this modification (from a 0 MeV baseline) is about 300.
The need for an energy scale is shown even more directly 
in the results of Table~\ref{T1}, which is a good summary of the 
benefits of the corrections. The $\chi^2$ drops after the improvements have  been employed, especially with the neutron projectile observables
at low energy. As the energy of the neutron projectile increases the changes become less and less important. The proton projectile
does not follow these trends at low energies because the coulomb force dominates. 
There were calculation error-bars given to 
each calculated quantity such that the error at forward angles was larger than backward angles  and the weighting of a total cross-section point was worth
about one third of a complete differential data set (measured from 5$^o$ to 174$^o$ at low energies). 
If one was to convert the results of Table 1
to a reduced
$\chi^2$, a $\chi^2$ of about 30 would be equivalent to a reduced $\chi^2$ of 1 in that every theoretical calculation point is within the error
bars chosen by the researchers which were meant to imitate typical experimental errors, so good results indeed as many of
the results of Table~\ref{T1} approach 30. For example a projectile energy of 35 MeV has very typical results, if one follows the results
at only 35 MeV in the figures below (Figs.~\ref{figds}-\ref{figtotreact}) one will see good improvement in every observable.
We also believe that the quality of the improvements will work
when the neutron projectile energy is less than 1 MeV because the energy dependence at extreme low energies is near constant and the imaginary
surface term, which should be small is reduced even smaller. Again all target nuclei at all projectile energies up to 200 MeV
benefitted from these adjustments as Fig.~1 depicts. Without the alterations
employed it can be seen that the $\chi^2$ is less than 50 for projectile energies of 200 MeV.

A word about
the coupled-channel framework which is already familiar 
to many~\cite{THOMPSON1988167, Raynal:94,thompson_nunes_2009}.
If the nucleus is non-spherical and has multi-pole deformations that
are either causing an excited rotation and / or vibrational collective mode then we can calculate these collective excited observables.
It starts by expanding the radius assuming a series of deformations
\begin{equation}
R(\theta) = R_0\left [1 + \beta_2Y_2^0(\theta) + \beta_4Y_4^0(\theta) + \beta_62Y_6^0(\theta)+ \ldots \right ],
\end{equation}
where the $\beta_\lambda$'s are deformation parameters. In a rotational model these are
corresponding to an quadrapole, hexapole, octupole rotational mode, where by symmetry,
only even modes are allowed~\cite{jelley1990fundamentals,Raynal:94}.
There is a partial wave dependent
asymmetry in the polar angle separating $\vec{r}$ and $\vec{r^\prime}$ in coordinate space. The same
common multi-pole expansion can be performed on the optical potential, with $x=\cos(\theta)$
\begin{equation}
U_\lambda(|\vec{r}|,|\vec{r^\prime}|) = \frac{1}{2}\int_{-1}^{+1}U(\vec{r},\vec{r^\prime,x}P_\lambda(x)\;dx.
\end{equation}
where $R = \vec{r^\prime} - \vec{r}$. 
The vibrational mode looks similar:
\begin{equation}
R(\theta) = R_0\left [1 + \sum_{\lambda} \alpha_\lambda \right ],
\end{equation}
where
\begin{equation}
\alpha_\lambda^\mu = \frac{\beta_\lambda^\mu}{\sqrt{2\lambda+1}}\left \{ (b_{\lambda,\mu} + (-1)^\mu b^+_{\lambda,-\mu} \right \},
\end{equation}
the $\mu$ stands for the magnetic quantum number and can be summed over to produce $\alpha_\lambda$.
As can be seen the
$\alpha_\lambda$ contain deformation parameters ($\beta_\lambda$), important for this work,
 and photon creation($b^+_{\lambda,-\mu} $) and destruction ($b_{\lambda,\mu}$) operators. With a vibrational
mode the phonons can follow all allowed electric and magnetic transitions.
One can then expand the potential in a Taylor series to first order as:
\begin{equation}
U(r,R) = U(r,R_0) + \frac{d}{dR_0} U(r,R_0) R_0 \sum_{\lambda}\alpha_\lambda Y_\lambda(\theta,\phi).
\label{vibrator}
\end{equation}
So these phonon excitations are derived from the non-spherical nature of the nucleus.
The vibrational, rotation, and a combination of both models can be calculated by common
distorted born approximation codes~\cite{ECIS06,FRESCO}. These codes take as input a
optical potential and then place it into the chosen vibrational/rotational mode.

We assumed that each nuclei was either distorted using either a simple first-order 
vibrational or rotational model. We stuck to a convention that if $A<130$ it was vibrational, otherwise rotational 
(we loosen these restrictions later) with the caveat that
if it was an even-odd or odd-even nucleus we always used vibrational because we had difficulty in getting ECIS to produce results 
otherwise (see Ref.~\cite{PhysRevC.93.044327} for a similar issue with ECIS studying
even-odd Calcium isotopes). For each target we chose  the deformation parameters, $\beta_\lambda$ 
from a variety of sources. We tried to rely on research that stemmed
from an inelastic experiment but in their absence we used deformation parameters created from a theoretical framework,
especially Ref.~\cite{MOLLER20161}. By using a plethora of sources for the distortion parameters we hope that the 
algorithm is robust. Once the initial research was finished we randomized the deformation parameter values in both directions multiple times, 
up to 30\%, to see
if the quality of the fits changed; they did not -- the statistical analysis always produced similar improvements. 
In general, as opined earlier, the larger the deformation, the larger the modification,
but besides that dependence the algorithm was stable. 
We used a deformed coulomb potential (with the same deformation parameter
as every other term). When the potential was effectively only a coulomb force (when the proton
could not penetrate the coulomb barrier at 1 MeV and at 5 MeV for large targets, see Table 1) the adjustments did very little which infers,
as expected, that the long range coulomb was not sensitive to this research. For each nucleus we made it a point to include at least 
three excited states or coupled channels.
In a later section we will study two nuclei where we will change the theoretical model and the reactions which are coupled 
to study the effect of this prescription on
these variances.

\begin{figure*}\includegraphics[width=6.95in]{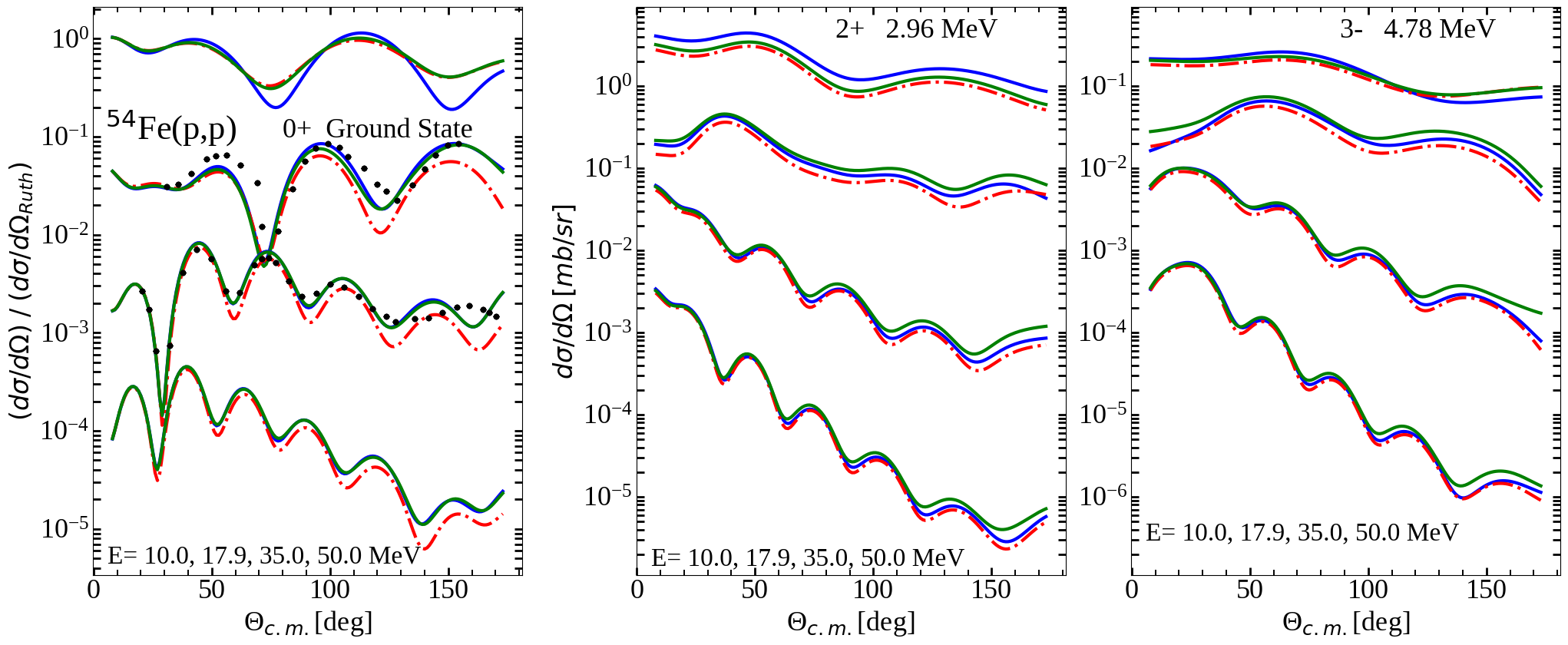}
	\caption{Differential cross-sections for a proton on $^{54}$Fe. 
	The legend and description is the same as Fig.~\ref{figds}. Note that the theoretical modification is poor at 10 MeV projectile energy, perhaps because the short term coulomb is important
	and it has not been modified, for more
        analysis see the text. For aesthetics the  E=17.9 MeV cross-sections have been reduced by a factor of 20, the E= 35.0 MeV by a factor of 400, the E = 50.0 by a factor of 8000.
	The experimental data at E = 17.9 MeV is from Ref.~\cite{GRAY1965565}, the experimental data from 35.0 MeV is from Ref.~\cite{O18p42rc, PhysRevC.21.844}
	For $^{54}$Fe $\beta_2 = 0.24$, $\beta_3 = 0.16$.}
\end{figure*}

\begin{figure*}\includegraphics[width=6.95in]{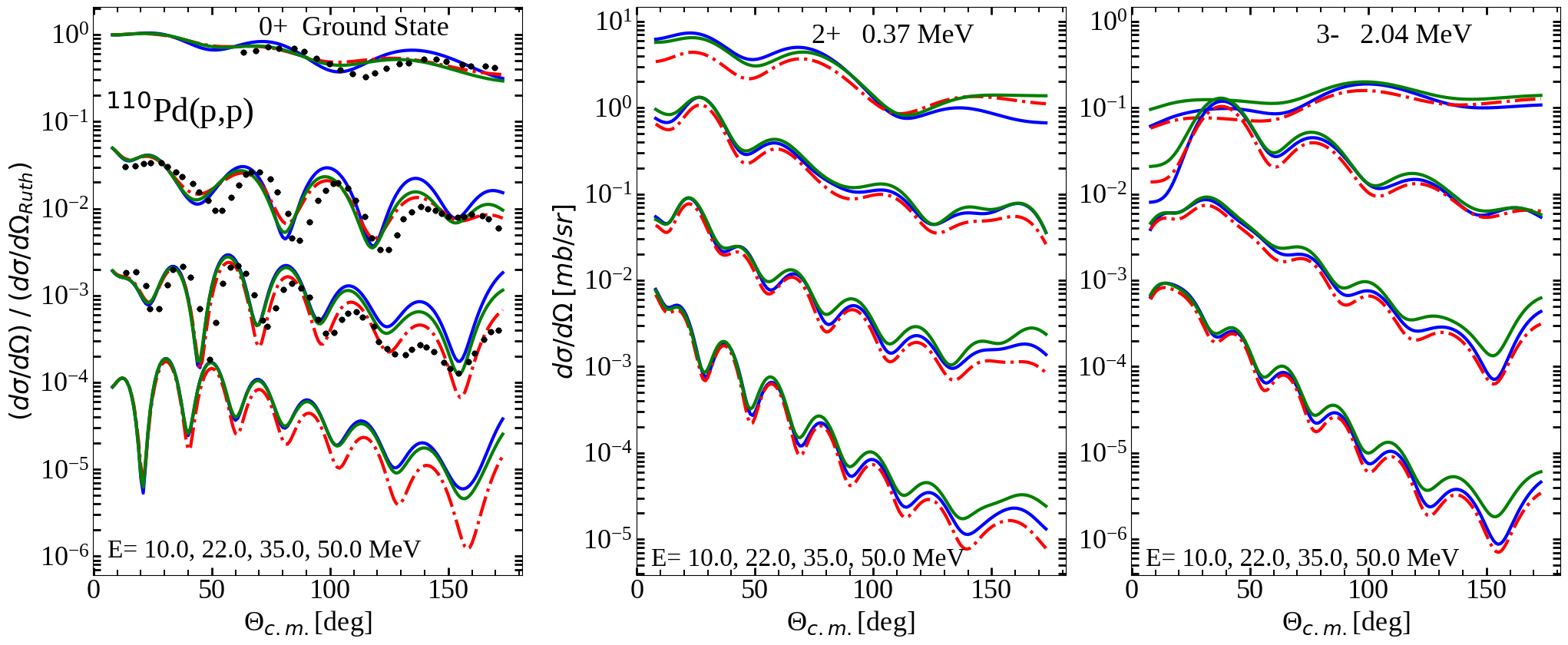}
\caption{Differential cross-sections for a proton on $^{110}$Pd. 
        The legend and description is the same as Fig.~\ref{figds}. Note that the theoretical modification 
	is poor at 10 MeV projectile energy, perhaps because the short term coulomb is important
        and it has not been modified, for more
        analysis see the text. For aesthetics the E=22.0 MeV cross-sections have been reduced by a factor of 20, the E= 35.0 MeV by a factor of 400, the E = 50.0 by a factor of 8000.
	The experimental data at E = 22 MeV and 35 MeV is from Ref.~\cite{Cereda:1982zz}. For $^{110}$Pd $\beta_2 = 0.24$, $\beta_3 = 0.11$}
\end{figure*}

\begin{figure*}\includegraphics[width=6.95in]{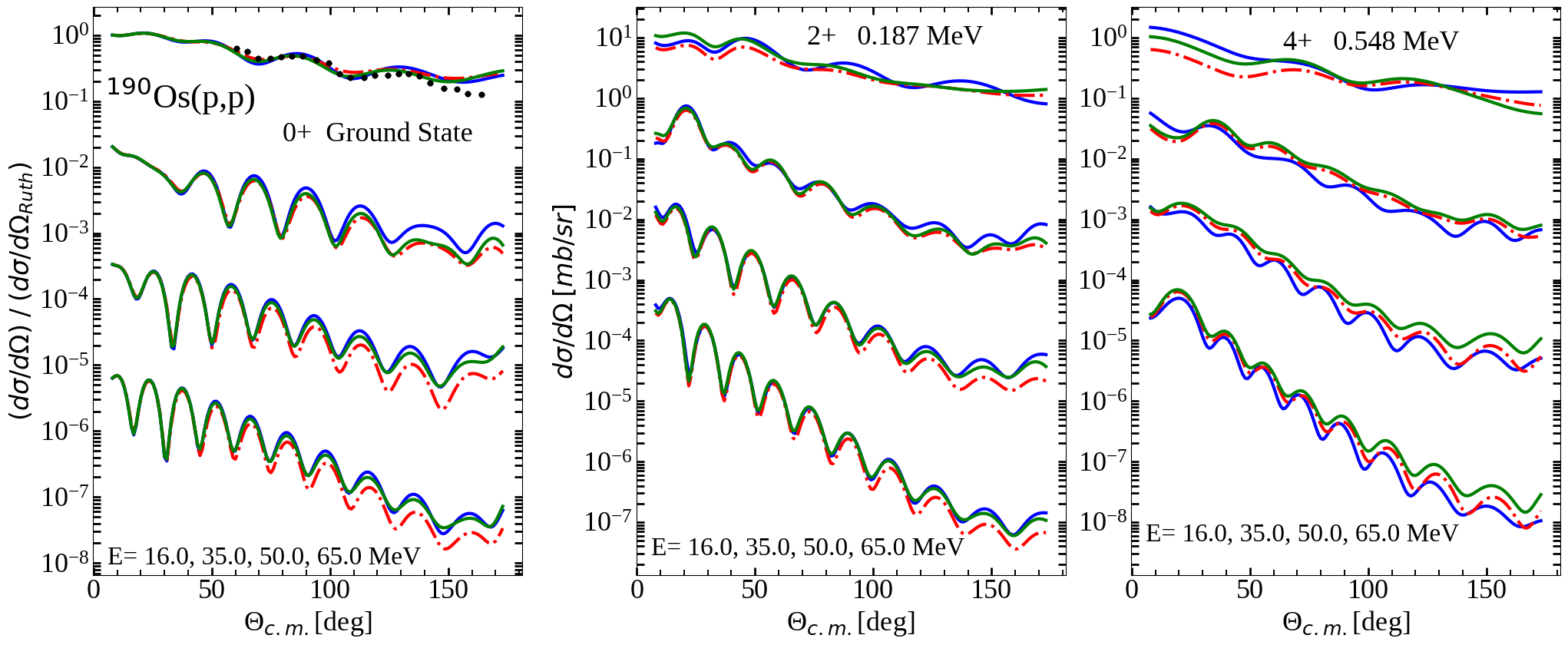}
\caption{Differential cross-sections for a proton on $^{190}$Os. The legend and description is the same as Fig.~\ref{figds},
        for more 
        analysis see the text. For aesthetics the E=35.0 MeV cross-sections have been reduced by a factor of 50, 
	the E= 50.0 MeV by a factor of 2500, the E = 65.0 by a factor of 125000.
        The experimental data at E = 16 MeV is from Ref.~\cite{KRUSE1971177}.For $^{190}$Os $\beta_2 = .187$, $\beta_4 = -.086$ }
\end{figure*}

As stated we had over a thousand cases, which meant over three-thousand different observables were tested. 
Since each case was independent, the fit was done in parallel, and the fitting procedure was done within the {\ it ECIS} software which
was the workhorse of this research. After getting a best fit for the real and imaginary surface terms from the code
was run again with those parameters as input and the fitting process was done again. This step was repeated, usually 5-7 times,
until convergence was reached. Overall the results were very consistent, the real surface term was increased from 0-6 MeV
and the imaginary surface term was decreased by 0-6 MeV. To summarize, (1) the real central volume term was decreased in an
analytical procedure, (2) the real central, surface, and spin-orbit radii  were decreased in an analytical procedure
(3) The real and imaginary surface amplitudes were lastly fit using {\it ECIS}.

We chose four nuclei for most of our
representations in this work: $^{12}$C,$^{54}$Fe, $^{110}$Pd (all treated as a vibrator), and $^{190}$Os (treated as a rotator), 
a small subset but representative of
the data as a whole. All results with $^{12}C$ used the potential from Ref.~\cite{PhysRevC.80.034608}, the 
rest used the optical potential of Ref.~\cite{KONING2003231}. A good result for our solid green line modification is when it
coincides often with the original solid blue line calculation {\bf on the left-most elastic observables 
graph only}. The reason for the singling out of the first panel is that the 
optical potentials were fit only to the elastic scattering observables of the left panel. The two right panels represent
usually the two lowest excited energy levels of the target nucleon. They are an integral part of our discussion but they were 
not used as part of the fitting routine.

Some systematic results that can be gleaned from these eight figures of differential cross-sections(\ref{figds}-\ref{figendds}). 
As already discussed,
at low energies the effects are more pronounced, thus here we displayed nothing over an energy of 65 MeV even though
we calculated observables up to 200 MeV. Generally, our modification to the potentials returns the optical
potential to close to its original status (green line coincides often with blue line in the leftmost panel), the largest effects and the
largest $\chi^2$ happens at low projectile energy. 
The right two panels of each figure deserve mention. These represent the fit to
the 2+ and 3- states. As stated earlier they were not used in developing the altercations however systematically what it
shows in the differential cross-sections is that there is a distinct difference
in the sensitivities of the deformation parameter between a coupled-channel calculation and a non coupled-channel
calculation. Generally there have been large 
uncertainties on this deformation parameter (as an $^{16}O$ example see Ref.~\cite{Svenne:2016fzb}) and
generally strict theoretical calculations of this deformation parameter under predict their
experimental counterparts (compare the $^{16}O$ 
experimental Ref.~\cite{Svenne:2016fzb} to the theoretical Ref.~\cite{MOLLER20161}). Our results also show that
there is an increased sensitivity to the parameter if our modified coupled-channel {\bf is} used. Our corrected green solid line
in the two rightmost panels of the differential cross-sections is often the calculation with the 
largest magnitude, (Figs.~\ref{figds}-\ref{figendds})
thus showing a heightened influence of the deformation parameter ($\beta_i$);
a smaller deformation value is needed to achieve fitting experimental observables.
This coincides with the coupled channel calculation of Ref.~\cite{Svenne:2016fzb} and it
was also found by authors involved in this research ( {\it  Ranga et. al. submitted to {\it Journal of Physics G}}).

Proton scattering at low energies is denominated by the coulomb force and thus our prescription has little effect at projectile 
energies which are approximately
equivalent to or less than the couomb barrier. Since we do not adjust the coulomb parameter this is expected. It is also important to note
that all these calculation except for carbon use Koning and
Delaroche~\cite{KONING2003231} global optical potential and sometimes the original fit is good (as for $^54$Fe) and at other times
the quality is less than desired (as for $^110$Pd). This prescription was built to duplicate the original theoretical results so
its validity is strongly correlatesd to the quality of the original fit. Carbon calculations used Refs.~\cite{PhysRevC.80.034608, PhysRevC.89.049904}
and the same statement is true. The fits for $^{12}$C varied and thus it is important to ascertain the quality of your original
global potential before applying this prescription. This will become even more apparent with a neutron projectile (see Fig.~\ref{figdsn}).

\begin{figure*}\includegraphics[width=6.95in]{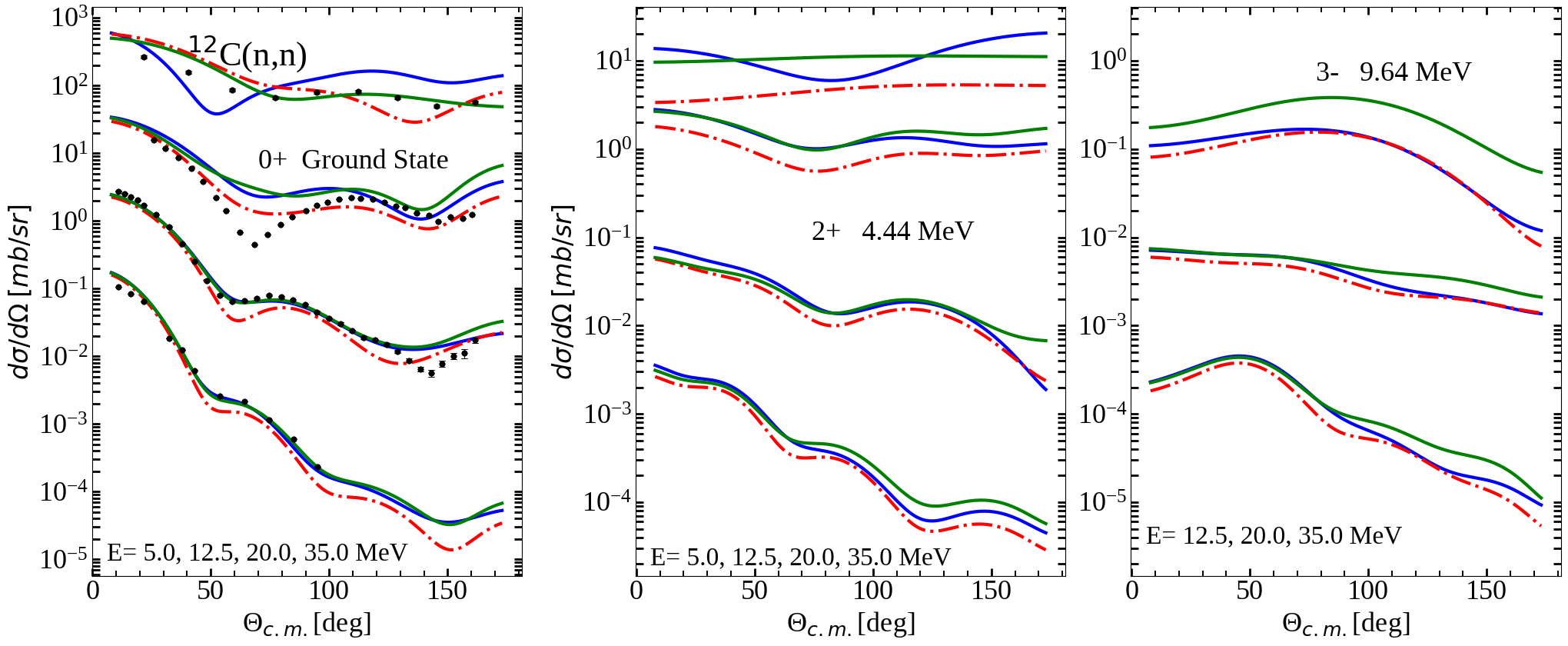}
\caption{Differential cross-sections for a neutron on $^{12}$C. In all calculations the solid blue line is a calculation using the original spherical
        DWBA potential, the solid green line has modified that potential and uses a coupled-channel calculation. The red dashed line is the
        inconsistent hybrid of the original potential used in a coupled-channel environment. The left panel represents the elastic (ground state only)
        differential cross-section.
        for four different neutron energies, lowest energy at the top, highest energy at the bottom. this left panel presents one of the observables
        which we fit to, a good fit would be when the solid green calculation is close to the original solid blue calculation. In contrast to a
	proton projectile these neutron projectile calculations show more sensitivity to the modification and we are also able to show 
	lower projectile energies due to the absence of a coulomb barrier. The central and right
        panel represent inelastic differential cross-sections which were not used in the fittings, they however show the sensitivity of the modification
        to the collective 2+ and 3- inelastic cross-sections, again the projectile energies are the lowest at the top and highest at the bottom. For more
        analysis see the text.
        For aesthetics the  E=12.5 MeV cross-sections have been reduced by a factor of 20, the E= 20.0 MeV by a factor of 400, the E = 35.0 by a factor of 8000.
	The experimental data at E = 5.0 MeV is from Ref.~\cite{WHITE198013}, the data from 12.5 MeV is from Ref.~\cite{PhysRevC.28.2212}, 
	the data from 20.0 MeV is from Ref.~\cite{OLSSON1989505}, and the data from 35.0 MeV is from Ref.~\cite{NIIZEKI1990455}.}
\label{figdsn}\end{figure*}

\begin{figure*}\includegraphics[width=6.95in]{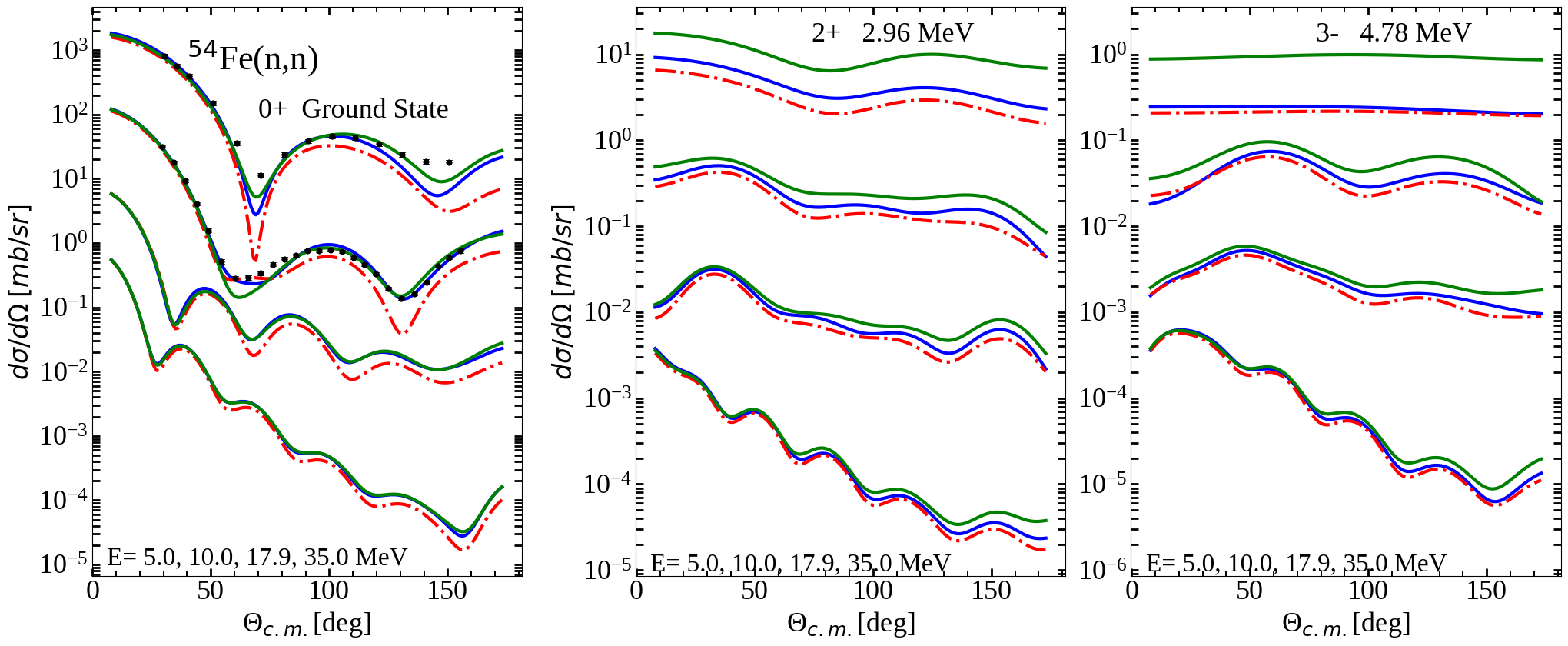}
\caption{Differential cross-sections for a neutron on $^{54}$Fe. The legend and description is the same as Fig.~\ref{figdsn},
        for more
        analysis see the text.
	For aesthetics the  E=10.0 MeV cross-sections have been reduced by a factor of 20, the E= 17.9 MeV by a factor of 400, the E = 35.0 by a factor of 8000.
	The experimental data from E = 5.0 MeV projectile energy is from Ref.~\cite{VANHOY2018107}, the data for E = 10.0 MeV is from Ref.~\cite{ELKADI1982509}.}
\end{figure*}

\begin{figure*}\includegraphics[width=6.95in]{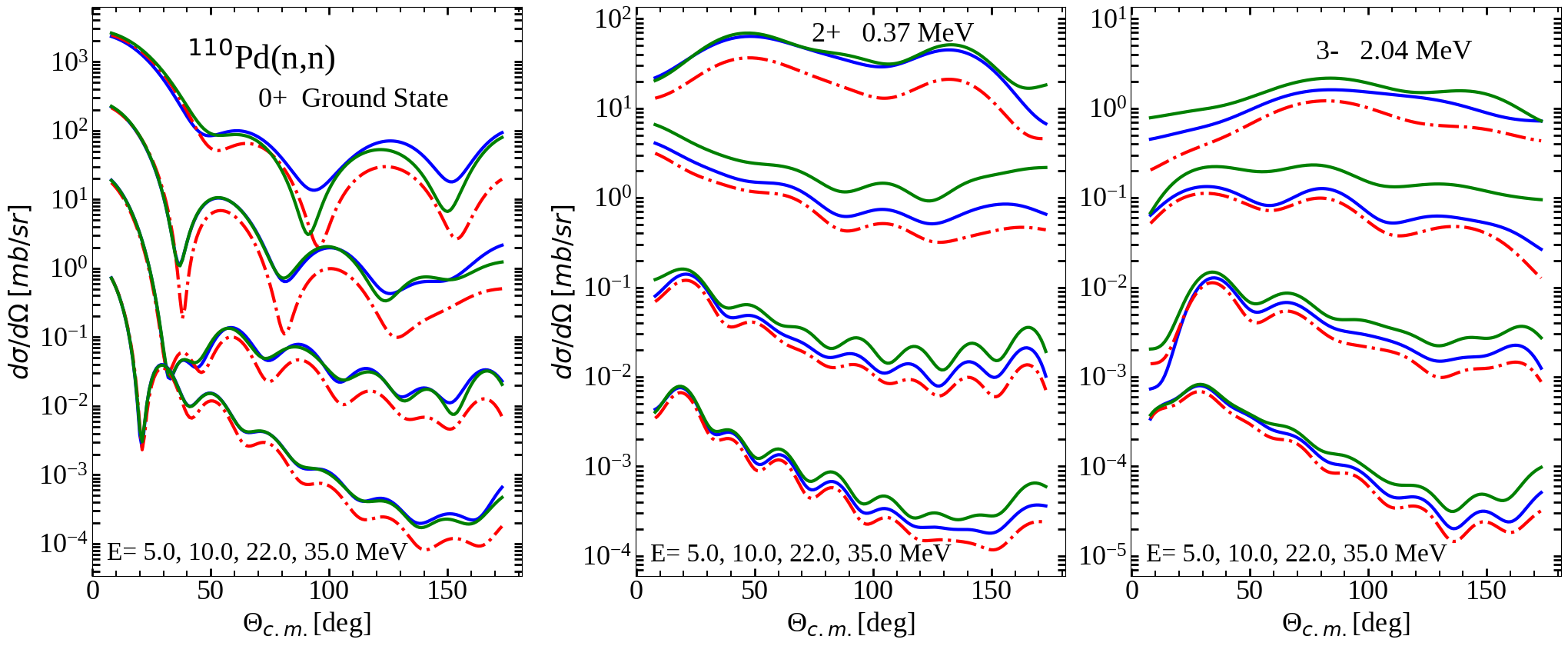}
\caption{Differential cross-sections for a neutron on $^{110}$Pd. The legend and description is the same as Fig.~\ref{figdsn},
        for more
        analysis see the text.
	For aesthetics the  E=10.0 MeV cross-sections have been reduced by a factor of 20, the E= 20.0 MeV by a factor of 400, the E = 35.0 by a factor of 8000.}
\end{figure*}

\begin{figure*}\includegraphics[width=6.95in]{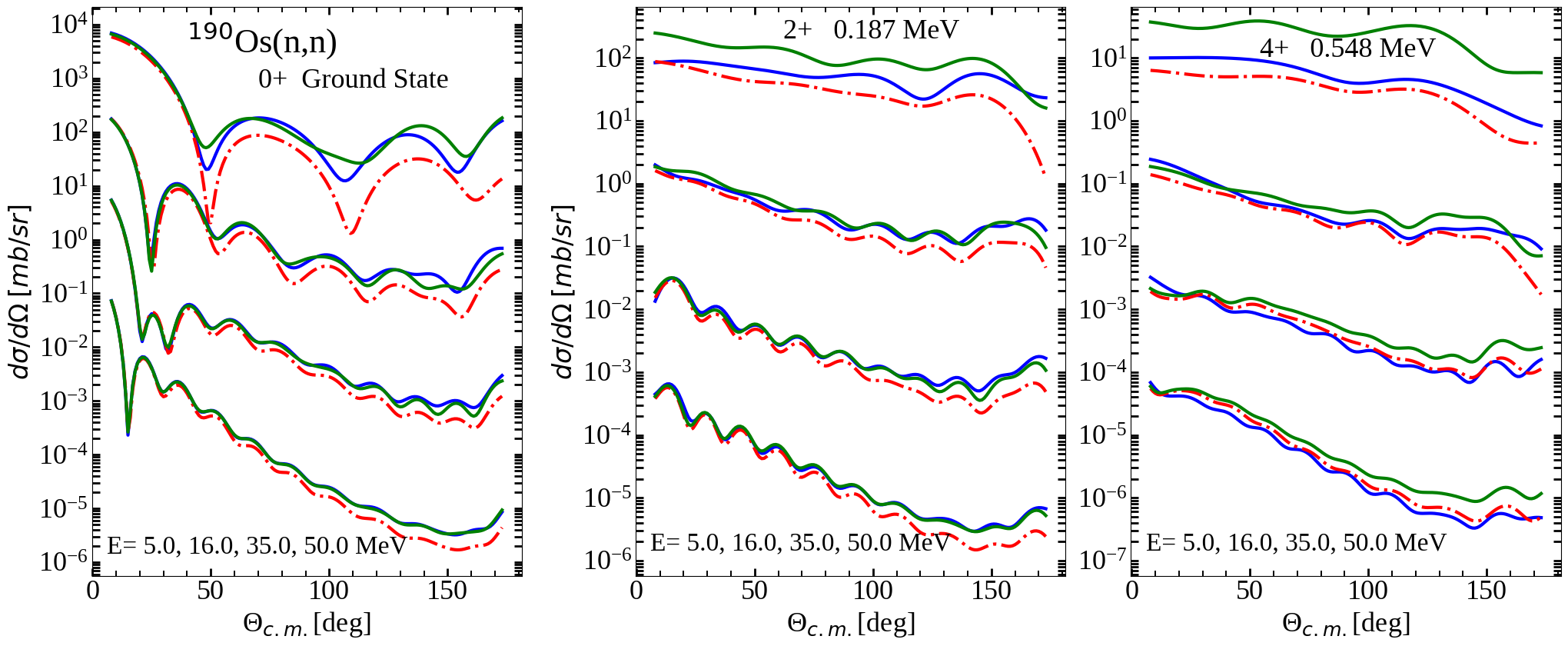}
\caption{Differential cross-sections for a neutron on $^{190}$Os. The legend and description is the same as Fig.~\ref{figdsn},
        for more
        analysis see the text. For aesthetics the E=16.0 MeV cross-sections have been reduced by a factor of 50, 
        the E= 35.0 MeV by a factor of 2500, the E = 50.0 by a factor of $1.25E5$.}
\label{figendds}\end{figure*}

Some systematic results that can be gleamed from the four figures of polarization graphs(\ref{figpol}-\ref{figendpol}).
As already discussed,
at low energies the effects are more pronounced, thus here we displayed nothing over an energy of 65 MeV even though
we calculated observables to 200 MeV. Generally, our modification to the potentials returns the optical
potential to its original status (green line coincides often with blue line on the leftmost panel) 
even though the spin-orbit potential has
only been modified in the real term's radius
and the polarization was used in fitting but in our weighted $\chi^2$ it had
the smallest weighting.
The inelastic states (the two rightmost panels of Figs.~\ref{figpol}-\ref{figendpol}) also have systematic
characteristics in that here, decisively, the calculation of our modified coupled-channel result (green solid line)
coincides most often with the erroneous red-dashed line calculation which is a coupled-channel calculation
which uses an inconsistent, non coupled-channel developed, potential. This signifies to us that especially at
low energies, the coupled-channel aspect of the theory is an integral part of the calculation and that
our modification for the inelastic spin-observables plays a secondary role. 

\begin{figure*}\includegraphics[width=6.95in]{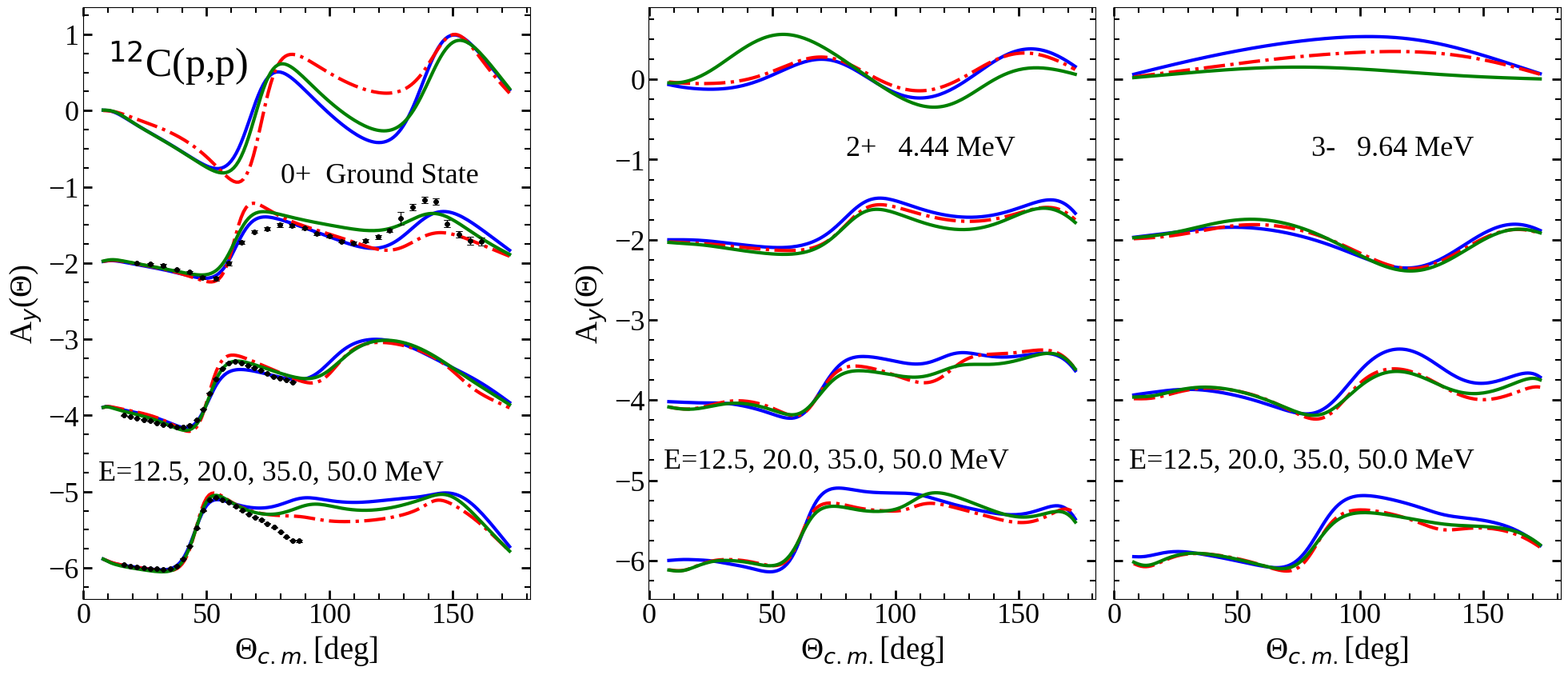}
\caption{Spin Polarizations for a proton on $^{12}$C. In all calculations the solid blue line is a calculation using the original spherical
        DWBA potential, the solid green line has modified that potential and uses a coupled-channel calculation. The red dashed line is the
        inconsistent hybrid of the original potential used in a coupled-channel environment. The left panel represents the elastic (ground state only)
	spin polarization ($A_y$) 
        for four different proton energies, lowest energy at the top, highest energy at the bottom. this left panel presents one of the observables
        which we fit to, a good fit would be when the solid green calculation is close to the original solid blue calculation. Note that the spin-orbit
	aspect of the potential was modified only in the real radius parameter and yet the fits are quite good. The central and right
        panel represent inelastic spin polarizations which were not used in the fittings, they however show the sensitivity of the modification
        to the collective 2+ and 3- inelastic cross-sections, again the projectile energies are the lowest at the top and highest at the bottom. For more
        analysis see the text.
        For aesthetics the  E=20.0 MeV polarizations have been offset by a factor of -2, the E= 35.0 MeV by a factor of -4, the E = 50.0 by a factor of -6.
	The experimental data at E = 20.0 MeV is from Ref.~\cite{CRAIG1966177}, at 35.0 MeV and 50.0 MeV is from Ref.~\cite{IEIRI1987253}}
\label{figpol}\end{figure*}

\begin{figure*}\includegraphics[width=6.95in]{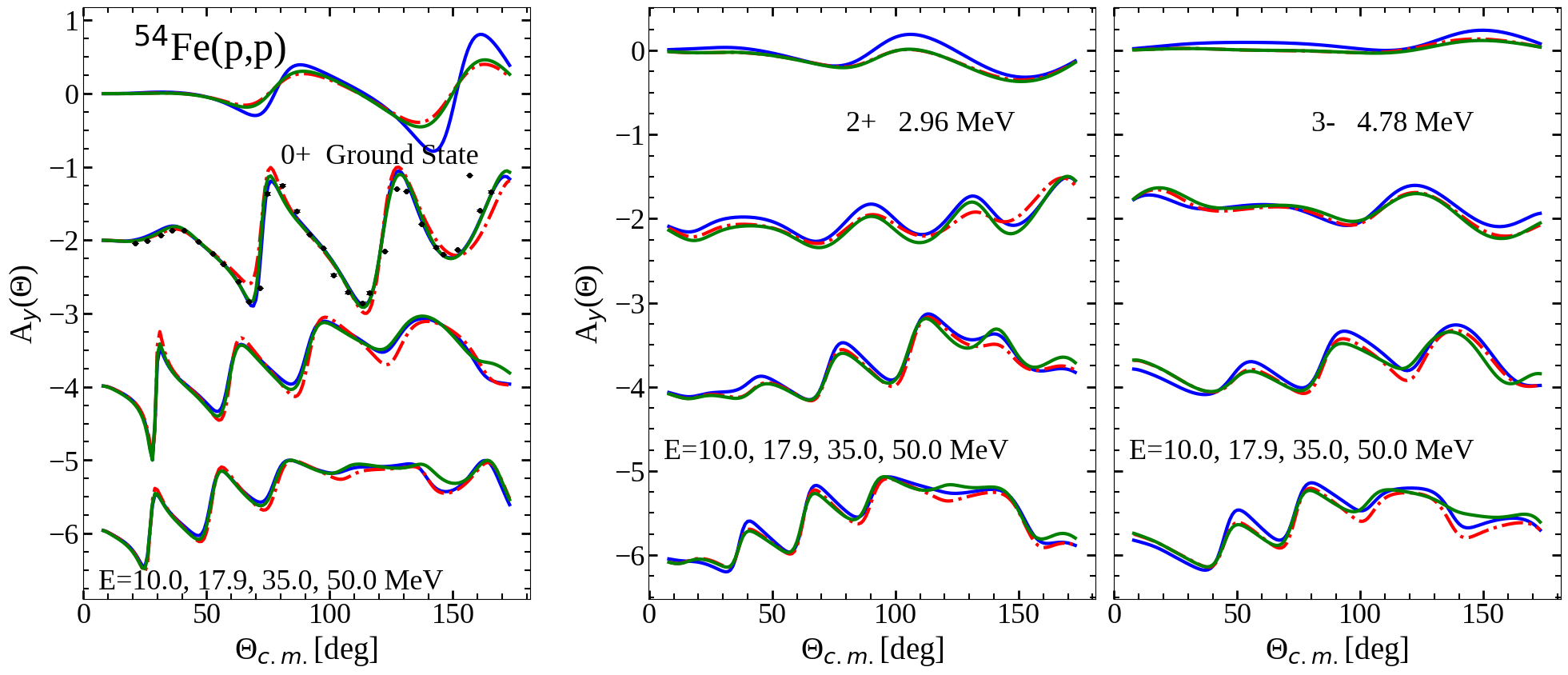}
\caption{Spin Polarizations for a proton on $^{54}$Fe. The legend and description is the same as Fig.~\ref{figpol},
        for more
        analysis see the text. For aesthetics the  E=17.9 MeV polarizations have been offset by a factor of -2, 
	the E= 35.0 MeV by a factor of -4, the E = 50.0 by a factor of -6.   The experimental data at E = 17.9 MeV is from Ref.~\cite{VANHALL197763}.}
\end{figure*}

\begin{figure*}\includegraphics[width=6.95in]{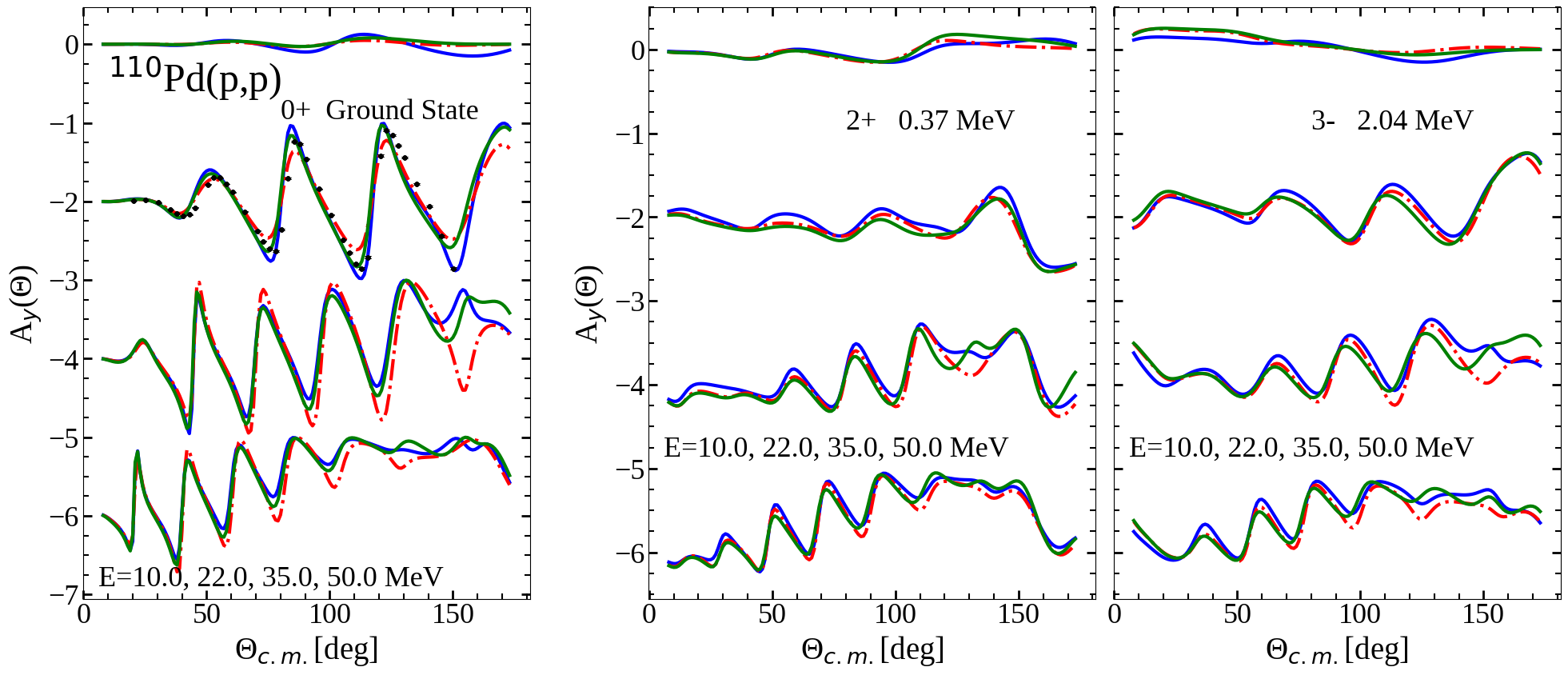}
\caption{Spin Polarizations for a proton on $^{110}$Pd. The legend and description is the same as Fig.~\ref{figpol},
        for more
        analysis see the text. For aesthetics the  E=22.0 MeV polarizations have been offset by a factor of -2, 
        the E= 35.0 MeV by a factor of -4, the E = 50.0 by a factor of -6.   The experimental data at E = 22.0 MeV is from Ref.~\cite{AOKI1983413}.}
\end{figure*}

\begin{figure*}\includegraphics[width=6.95in]{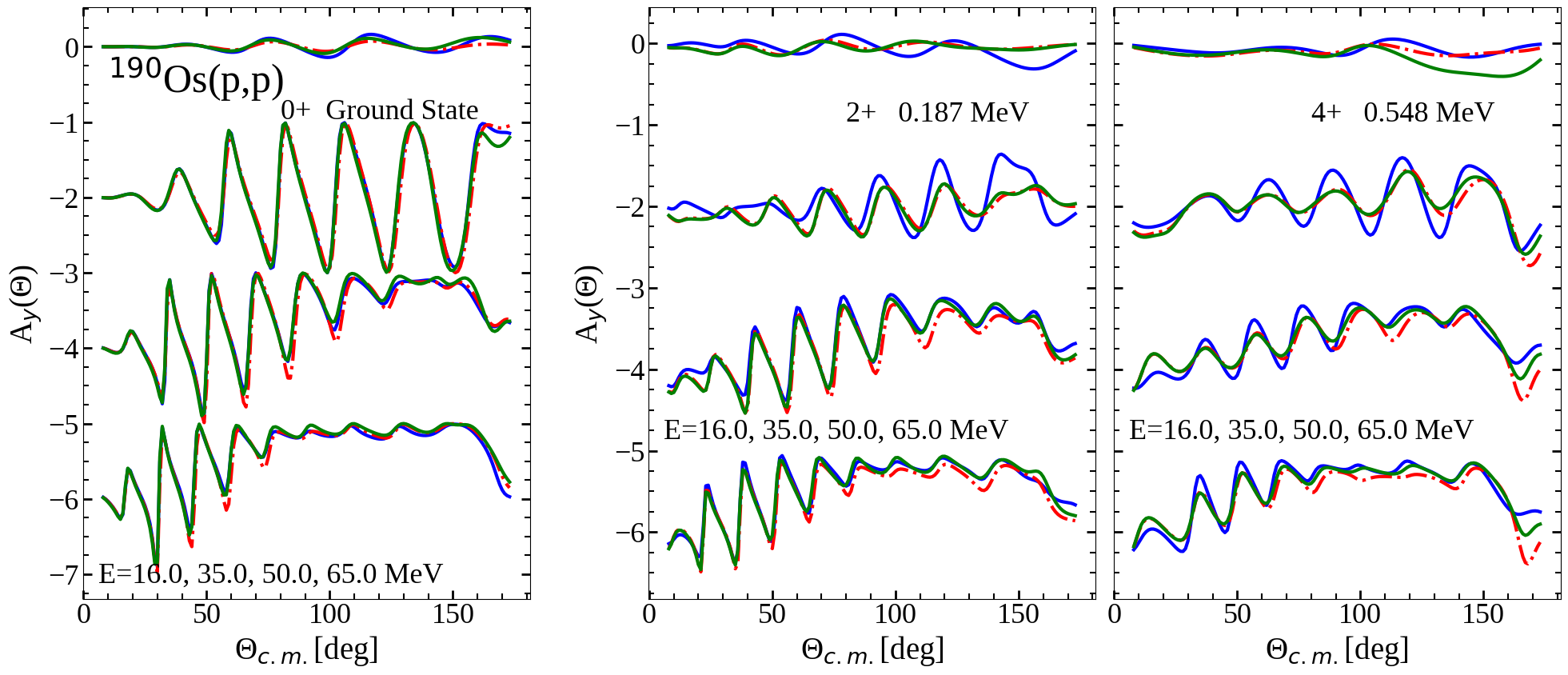}
\caption{Spin Polarizations for a proton on $^{190}$Os. The legend and description is the same as Fig.~\ref{figpol},
        for more
        analysis see the text. For aesthetics the  E=35.0 MeV polarizations have been offset by a factor of -2,
        the E= 50.0 MeV by a factor of -4, the E = 65.0 by a factor of -6. }
\label{figendpol}\end{figure*}

\begin{figure*}\includegraphics[width=6.95in]{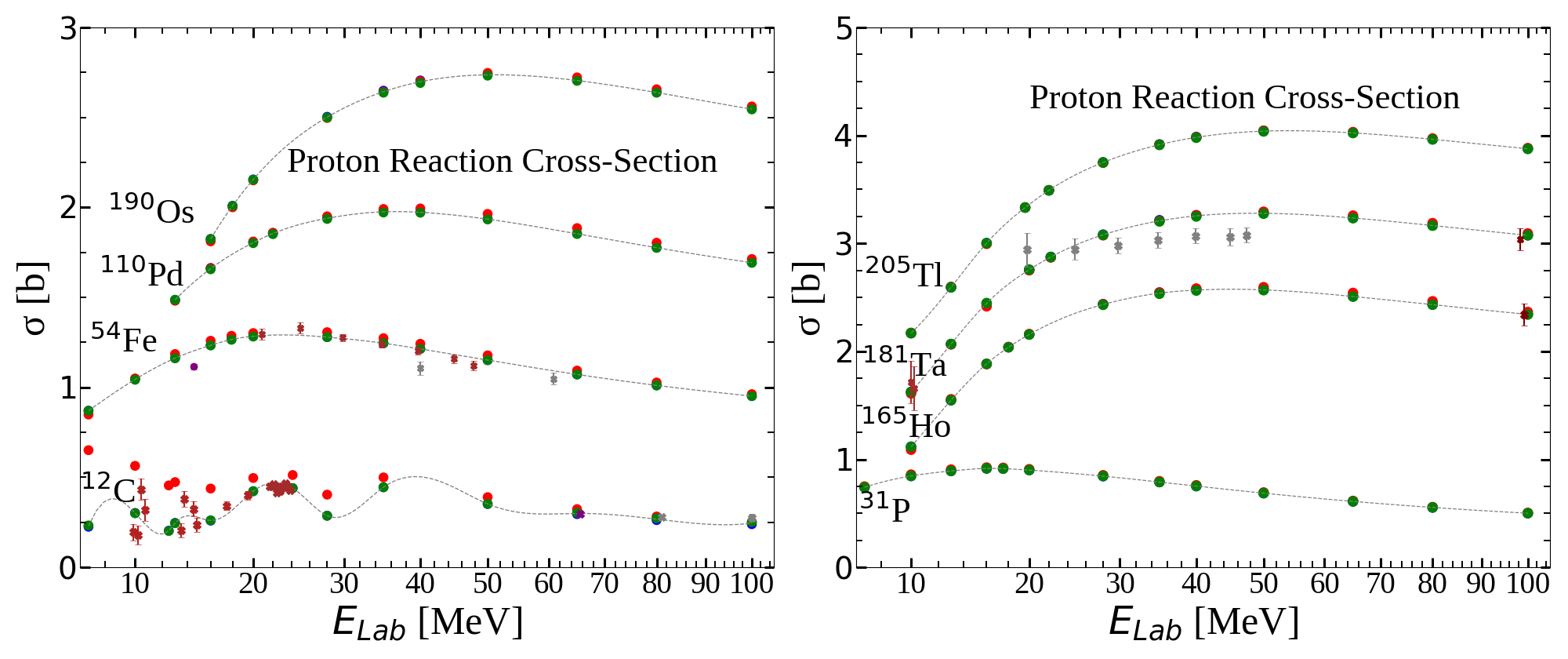}
	\caption{Eight different nuclei as targets, a proton as projectile and the total reaction (inelastic) cross-section is measured.
	In all calculations the solid blue circle is a calculation using the original spherical
        DWBA potential, the solid green circle has modified that potential and uses a coupled-channel calculation.
	The solid red circle is the
	inconsistent hybrid of the original potential used in a coupled-channel environment. The aim is to bring the modified potential (green circle) back to 
	the original potential (blue circle). If one cannot spot the original potential calculation it is because the improved potential (green circle) is
	on top of it which, in this figure, is true in all cases, the fit is excellent. 
	The thin dashed line goes through the average of the original and modified calculation to make it easy to follow the trend. 
	For more
        analysis see the text.
	In the left panel, for aesthetic reasons, 
	$^{54}$Fe is offset by 250 mb, $^{110}$Pd is offset by 500 mb, and $^{190}$Os is offset by 750 mb. In the right panel $^{165}$Ho is offset by 
	700 mb, $^{181}$Ta is offset by 1330 mb, and $^{205}$Tl is offset by 2010 mb.
        Experimental data for C12 comes from Refs.~\cite{PhysRevC.2.488,PhysRevC.12.1093,C12p142ra,INGEMARSSON1999341}, the experimental data
	for $^{54}$Fe comes from Refs.~\cite{doi:10.1139/p86-126,PhysRev.157.1001,PhysRevC.4.1114}, the 
	experimental data for $^{165}$Ho comes from Ref.~\cite{doi:10.1139/p66-155} and the experimental data for $^{181}$Ta comes from
	Refs.~\cite{doi:10.1139/p66-155, ABEGG1979109, PhysRev.129.2198}.
	}
\label{figproreact}\end{figure*}

Lastly we look at the total cross-sections. Here we add four odd A nuclei to our study ($^{31}P$, $^{165}Ho$, $^{185}Ta$, and $^{205}Tl$)
because, as we shall see, the differences for the total cross-sections can be more pronounced than the differential cross-sections,
more sensitivity is exhibited.
That said the proton reaction cross-section is barely affected by this modification. We assume that the
inelastic aspect of the coulomb force is therefore barely affected by the adjustments (this is independent
on the role of the coulomb in the dipole, quadrapole, and octupole deformations).
The neutron reaction cross-section does however show dramatic differences, especially at low energies. 
This quantity shows the error, probably the clearest of all the observables, in assuming a DWBA spherical global potential can function
in a coupled-channel environment (calculation
depicted by a red circle) which systematically runs high for low energy neutron scattering.
This implies that the undistorted potential inflates the inelastic part of the scattering amplitude.
Here we can really examine how our modification (green circle) pulls the theory back down
to the original non coupled-channel calculation (blue circle). If we break our corrections
into two parts: (1) the modification of the real volume amplitude and the real radius
(2) the modification of the surface. Under analysis part (1) does lower this magnitude but 
only doing about 20\% of the correction. The surface term adjustment is actually 
responsible for the bulk of the correction, we found this result to be true in all cases.
This suggests that the altering of the radius and real volume amplitude are there to temper the
change of the surface term. 
What we walk away understanding is that intrinsically we can mimic a distortion by shrinking
the imaginary thick surface while growing the real surface term and the neutron reaction
cross-section of Fig.~\ref{figneureact} then returns to its original value. To a lesser extent we see the same behavior
in the neutron total cross-section results of Fig.~\ref{figtotreact}. These
modifications are loosely based on theory but ultimately the algorithm chosen met the criteria
that it fit these total cross-sections effectively. This is not surprising because in our
weighted $\chi^2$ algorithm the total cross-sections were severely weighted.

The experimental results often do not coincide with either the original
optical potential or its coupled modification counterpart in Fig.~\ref{figtotreact}, that the 
fits are not ideal is common. It is difficult to fit
odd nuclei, small nuclei, and low energy neutron-nucleus scattering (below 20 MeV) with an optical model
ansatz. This is a reminder that the theoretical results do improve in these ranges; the adjustments do return the calculation
to near their original values even though the original potential has some shortcomings. We hope
that these corrections are independent of the potential and thus a good amount of work is still focused on
the building of better potentials which handle the many resonances and compound states  of low energy scattering (both
phenomenological and microscopic). Dispersive potentials have played an important role in low energy
scattering~\cite{DICKHOFF2019252}
 and we believe that these modifications are well suited for those calculations because of
only small changes to the imaginary terms.

A general statement can be made about our modifications looking at these total reaction
plots (Figs.~\ref{figproreact}-\ref{figtotreact}). Overall our improvements,
in a coupled-channel context, lower the reaction cross-section
for the nuclear force, raise the nuclear elastic cross-section (see Figs.~\ref{figds}-\ref{figendds}, 
and thus slightly raise the total nuclear cross-section all towards the original values. 
As always, at low projectile energy the 
modification is most important.

\begin{figure*}\includegraphics[width=6.95in]{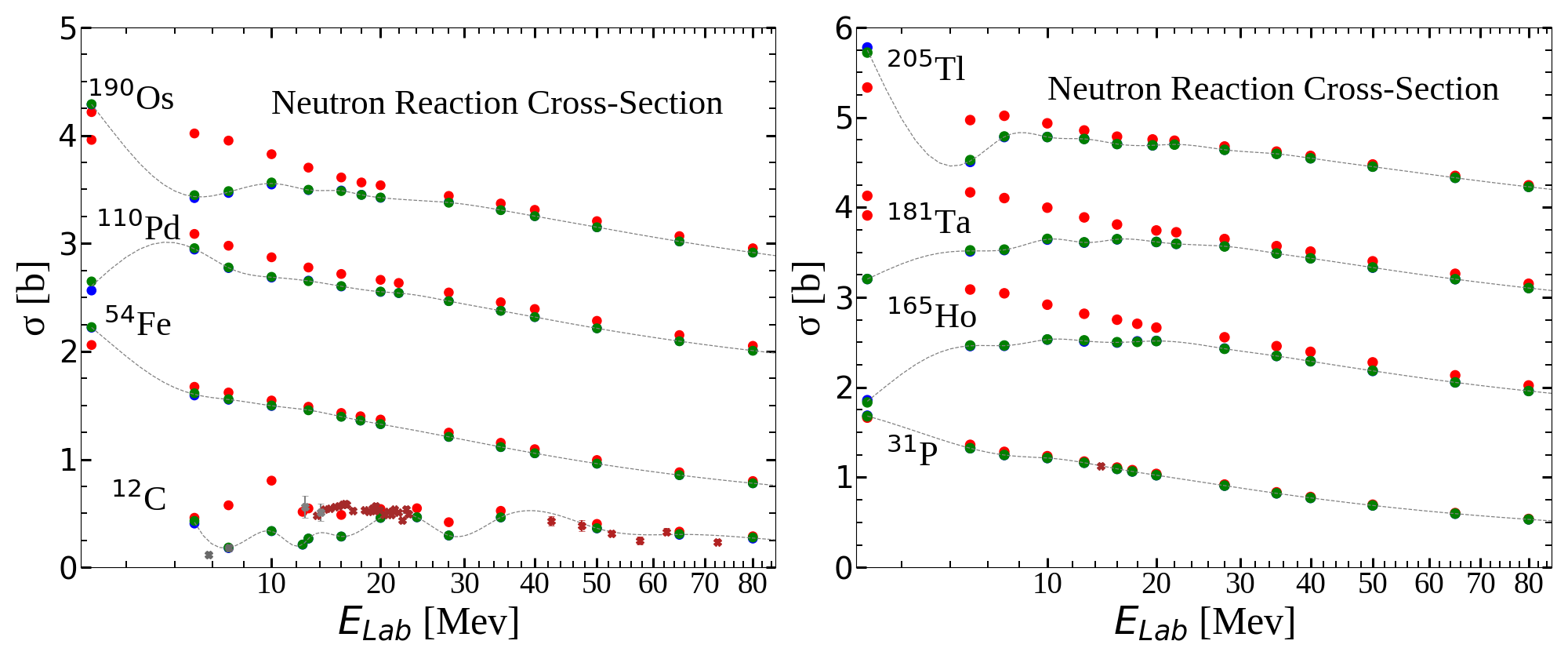}
\caption{Eight different nuclei as targets, a neutron as projectile and the total reaction (inelastic) cross-section is measured.
The legend and description is the same as Fig.~\ref{figproreact}. Note that this observable shows dramatic differences between the 
unmodified and the modified coupled-channel calculations,
        for more
        analysis see the text.
	In the left panel, for aesthetic reasons,
        $^{54}$Fe is offset by 20 mb, $^{110}$Pd is offset by 740 mb, and $^{190}$Os is offset by 960 mb. In the right panel $^{165}$Ho is offset by
        200 mb, $^{181}$Ta is offset by 1220 mb, and $^{205}$Tl is offset by 2190 mb.
	Experimental data for C12 comes from Refs.~\cite{doi:10.1080/00223131.2002.10875126,PhysRev.174.1147,PhysRev.100.174,PhysRev.112.486},
	the experimental data for $^{31}P$ comes from Ref.~\cite{Flerov1957InelasticCC}
}
\label{figneureact}\end{figure*}

\begin{figure*}\includegraphics[width=6.95in]{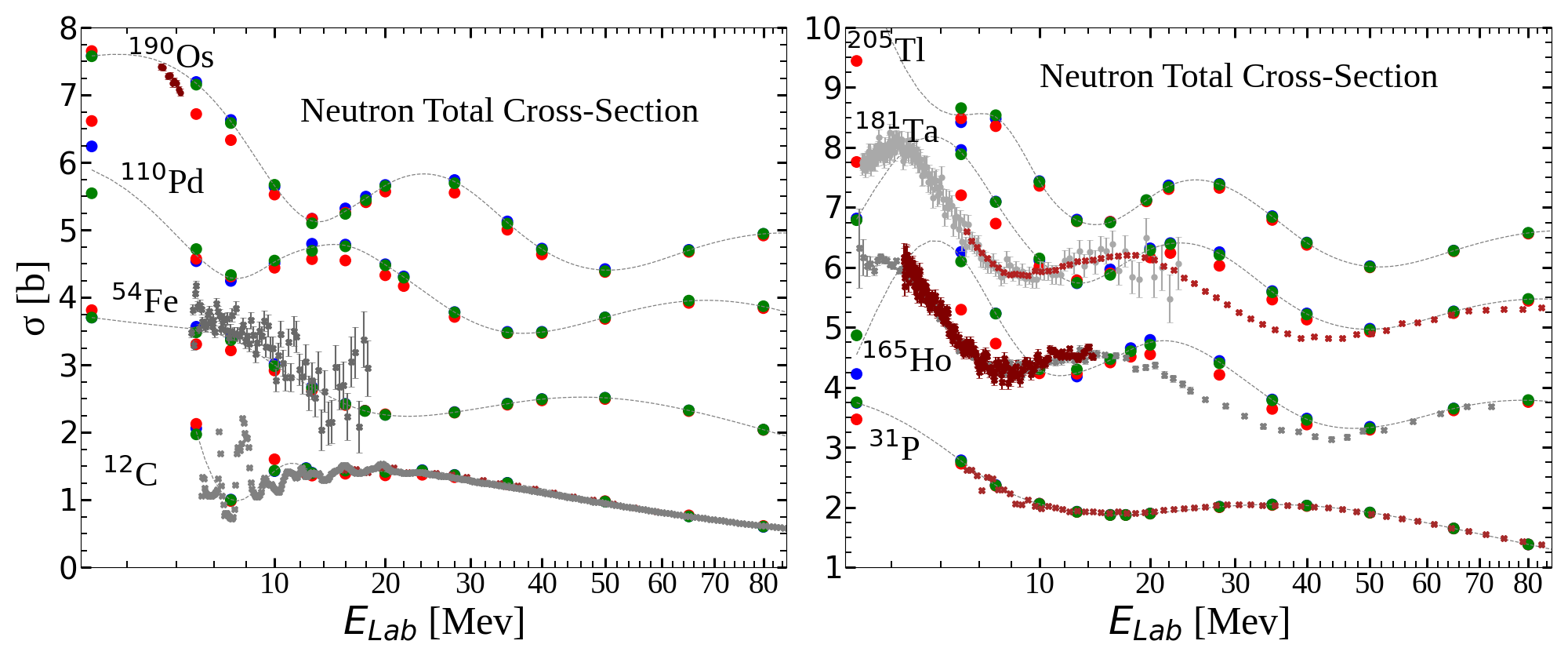}
\caption{Eight different nuclei as targets, a neutron as projectile and the total (elastic and  inelastic) cross-section is measured.
The legend and description is the same as Fig.~\ref{figproreact}. Note that this observable shows dramatic differences between the
unmodified and the modified coupled-channel calculations,
        for more
        analysis see the text.
        In the left panel, for aesthetic reasons,
        $^{110}$Pd is offset by 400 mb, and $^{190}$Os is offset by 200 mb. In the right panel $^{165}$Ho is offset by
        -700 mb, $^{181}$Ta is offset by 800 mb, and $^{205}$Tl is offset by 1700 mb.
	Experimental data for C12 comes from Refs.~\cite{SHANE2010468,PhysRevC.63.044608}, experimental data
	for $^{54}$Fe comes from Ref.~\cite{10.1007/978-94-009-7099-1_26}, and the experimental data for for $^{190}$Os comes
	from Ref.~\cite{thesis2}.In the right panel the experimental data for $^{165}$Ho comes from Refs.~\cite{PhysRevC.2.1862,PhysRevC.3.576},
	the experimental data for $^{181}$Ta comes from Refs.~\cite{doi:10.1080/00295639.2019.1570750, PhysRevC.47.237}.}
\label{figtotreact}\end{figure*}

\begin{figure}\includegraphics[width=3.48in]{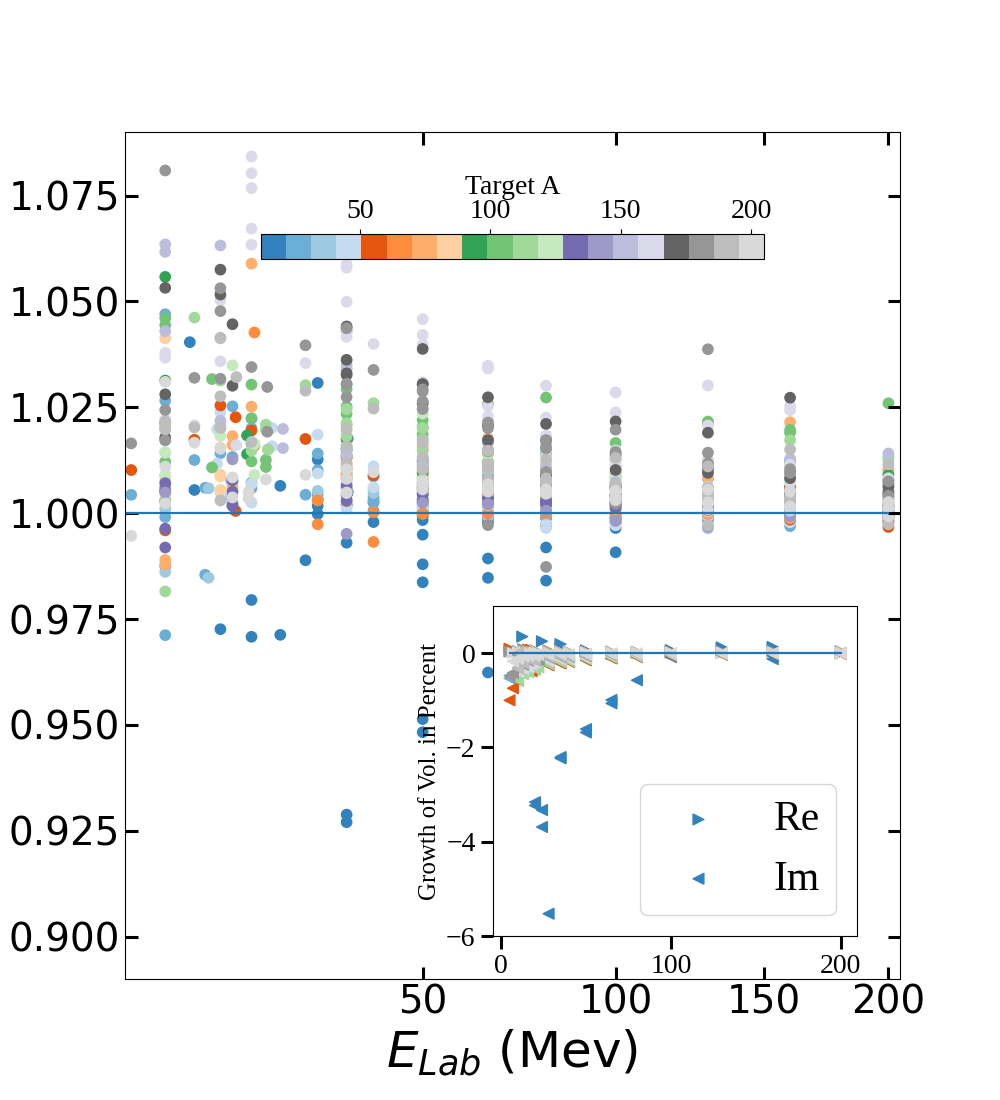}
	\caption{The main plot is a measure of how the magnitude (rms) 
	central volume changes (in percent) when the modification is applied, this contains over a thousand points.
The inset looks at a narrower set of nuclei ($^{12}$C, $^{31}$P,$^{54}$Fe,$^{110}$Pd,$^{165}$Ho,
	$^{181}$Ta,$^{190}$Os, and $^{205}$Tl), and shows how the imaginary volume shrank while the real volume grew.}
\label{figtovol}\end{figure}

Examining the change to the volume is illustrative as reflected in Fig.~\ref{figtovol}. More than 99.5\% of the time the real
volume of the central potential (including the surface term) grew by at most 5\%.
The only anomaly to this was the Carbon 12 nucleus at low energies. This was
caused, we believe, because of Carbon's extremely large distortion parameter combined with its
relatively small size caused it to be unsystematic in its behavior. Historically the smaller nuclei
have struggled to be fit in an optical potential ansatz. Overall, the volume stayed about the same
as depicted in Fig.~\ref{figtovol}. The real central volume grew by a few percent, the imaginary
central volume shrunk by a few percent. On average the magnitude of the central potential grew by 1\% and,
not depicted, the spin-orbit volume dropped by the same amount. As the figure depicts this was the
statistical average but there were outliers.  The inset of this figure shows details on our
illustrative subset of target nuclei (real versus imaginary). Because of these outliers we did
not make the prescription completely analytical.

The energy dependence of these optical potential terms has been studied~\cite{PhysRevC.93.064311} in the context of using
coupled channels. We also see a decrease in the energy dependence of the real central amplitude and the imaginary surface amplitude
but at the expense of adding a small artificial energy dependence to the real radius parameters. We applaud
ways to look for lessening the energy dependence of these parameters through dispersion and/or coupled-channel techniques.

\subsection{Changing the reaction and theoretical model}
This
method could also be applied to other DWBA local potentials which are not  in the form of Woods-Saxon functions where
the easiest approach would be to fit the optical potential to a Woods-Saxon basis
because there is a clearly
defined surface term in that space.
One of the authors (S.~P.~Weppner) has used multiple volume and
spin-orbit terms and one surface term to 
describe numerical microscopic local r-space potentials). There are also inelastic channels beyond collective excitations but they all involve
additional parameters which, like the deformation parameter $\beta_\lambda$, goes to zero when that channel is closed therefore these modifications
could be used in a similiar fashion with spectroscopic factors and transition potentials.

\section{Conclusions}
\label{sec:conclusions}

Overall we have found that if a modification is applied to a standard DWBA optical potential it then can be suitable for a 
coupled channel calculation which, most agree, contains relevant physics especially needed at the challenging
low energy regions of less than 25 MeV.
This modification was applied to a wide range of nuclei and energies and can be instantly applicable to one's research
by bringing to bear the full physical intuition of the coupled-channel approach within any form of the
DWBA spherical optical potentials including the uncertainty-qualified forms~\cite{PhysRevC.107.014602},
in
fact this research could be a first step leading towards a more statistical approach.
This
method could also be applied to other DWBA local potential
and it can be altered easily to benefit other inelastic processes (exchange, break-up, fission, fusion) and 
projectile scattering beyond the single nucleon all of which use the optical potential as an ingredient in their 
compound theory.
The coupled channel environment has also been applied to a microscopic framework (see Ref.~\cite{HAGINO2022103951} for a recent
review article). Likewise microscopic surface energy calculations are being studied in earnest~\cite{PhysRevC.83.034305,PhysRevC.99.044315}.
Phenomenological models, at their best, can help inform microscopic approaches and we hope the methods
outlined in this research can do the same.


\subsection{Acknowledgments}
	This research was started while one of the authors, S.~P.~.Weppner was on sabbatical at the
	{\it Tata Institute of Fundamental Research} in Mumbai. He thanks that institute and especially
	I. Mazumdar for
	their generous support.


\section*{References}

\providecommand{\newblock}{}

\end{document}